\documentclass[aps,prb,twocolumn,a4paper,showpacs,superscriptaddress]{revtex4}

\usepackage[dvipdfmx]{graphicx}
\usepackage{amsmath}
\usepackage{amssymb}

\usepackage{color}
\usepackage{ulem} 

\newcommand{\bvec}[1]{\mbox{\boldmath $#1$}}

\begin{document}
\title{Density-Matrix Renormalization Group Study of Extended\\ Kitaev-Heisenberg Model}

\author{Kazuya Shinjo}
\email{shinjo@yukawa.kyoto-u.ac.jp}
\affiliation{Yukawa Institute for Theoretical Physics, Kyoto University, Kyoto 606-8502, Japan}
\affiliation{Computational Condensed Matter Physics Laboratory, RIKEN, Saitama 351-0198, Japan}
\author{Shigetoshi Sota}
\affiliation{Computational Materials Science Research Team, RIKEN AICS, Hyogo 650-0047, Japan}
\author{Takami Tohyama}
\affiliation{Department of Applied Physics, Tokyo University of Science, Tokyo 125-8585, Japan}


\date{\today}
             

\begin{abstract}
We study an extended Kitaev-Heisenberg model including additional anisotropic couplings by using two-dimensional density-matrix renormalization group method. 
Calculating the gound-state energy, entanglement entropy, and spin-spin correlation functions, we make a phase diagram of the extended Kitaev-Heisenberg model around spin-liquid phase.
We find a zigzag antiferromagnetic phase, a ferromagnetic phase, a 120-degree antiferromagnetic phase, and two kinds of incommensurate phases around the Kitaev spin-liquid phase.
Furthermore, we study the entanglement spectrum of the model and find that entanglement levels in the Kitaev spin-liquid phase are degenerate forming pairs but those in the magnetically ordered phases are non-degenerate.
The Schmidt gap defined as the energy difference between the lowest two levels changes at the phase boundary adjacent to the Kitaev spin-liquid phase. 
However, we find that phase boundaries between magnetically ordered phases do not necessarily agree with the change of the Schmidt gap.
\end{abstract}
\maketitle


\begin{figure}[t]
\begin{center}
\includegraphics[width=18pc]{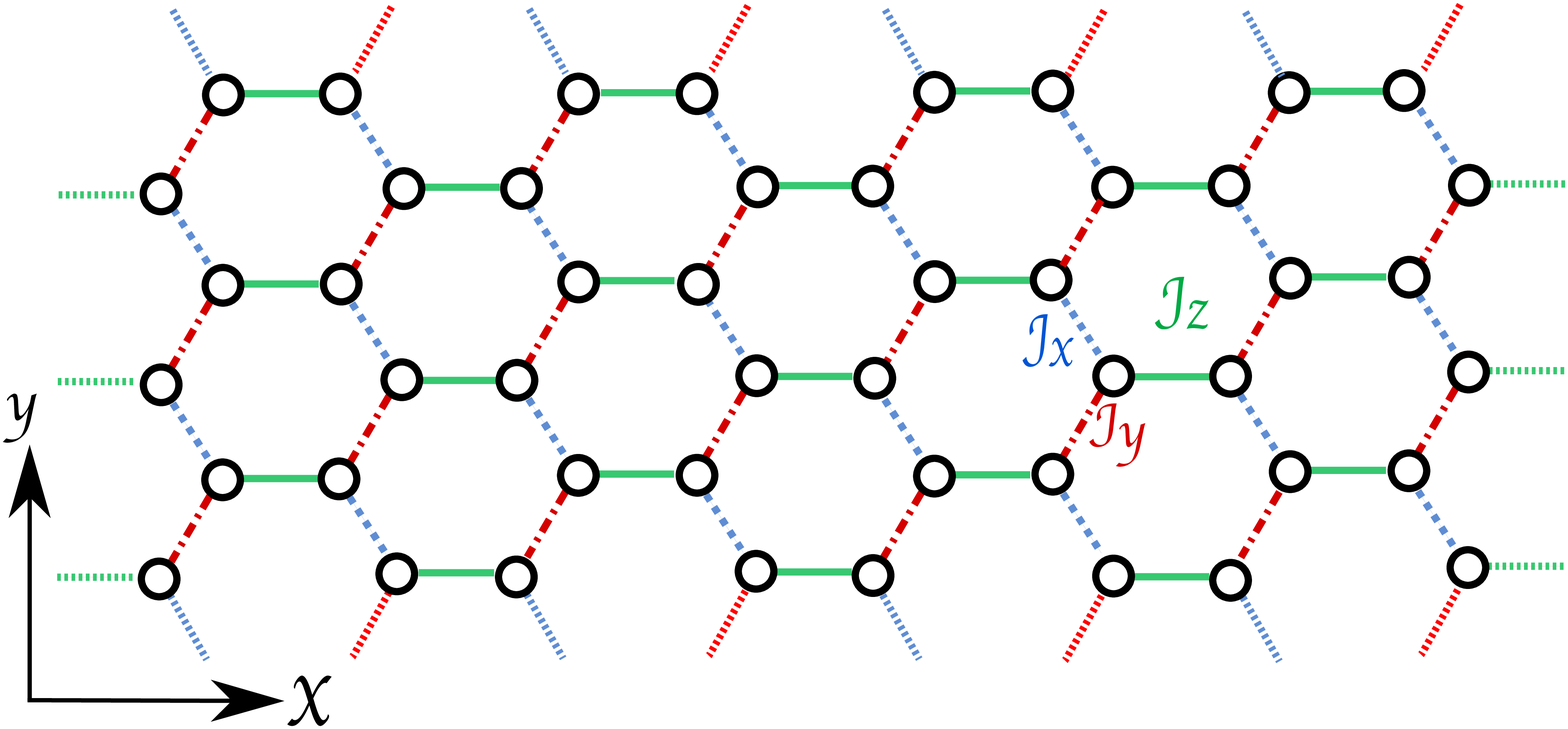}
\caption{(Color online) Honeycomb lattice with 6$\times$8 sites with periodic boundary condition. Blue dotted, red dashed-dotted, and green solid bonds labeled by $\mathcal{J}_x$, $\mathcal{J}_y$ and $\mathcal{J}_z$ have $S^xS^x$, $S^yS^y$ and $S^zS^z$ terms in a Kitaev model, respectively. We define the $x$-axis direction as an armchair-edge direction and the $y$-axis direction as a zigzag-edge direction.}
\label{honey}
\end{center}
\end{figure}

\section{Introduction}
Kitaev honeycomb lattice model is a spin-$1/2$ system on a honeycomb lattice.\cite{kitaev2006} The interactions between nearest neighbors are of $S^xS^x$, $S^yS^y$ or $S^zS^z$ type, depending on bonds $\mathcal{J}_x$, $\mathcal{J}_y$, and $\mathcal{J}_z$, respectively, as shown in Fig.~\ref{honey}.
The ground state of isotropic Kitaev model is known as a gapless Kitaev spin-liquid state characterized by gapless Majorana fermion excitations with two Dirac cones.\cite{knolle2014}
The spin-spin correlation of the gapless Kitaev spin-liquid is short-range, showing non-zero value only for the nearest-neighbor sites.\cite{baskaran2007}
However, perturbations such as antiferromagnetic (AFM) Heisenberg and Dzyaloskinski-Moriya interactions can qualitatively alter the nature of spin-spin correlation functions exhibiting a long-ranged power-law behavior.\cite{mandal2011}
 
Such a Kitaev-Heisenberg (KH) model has widely been studied as a prototype model for Na$_2$IrO$_3$ and its phase diagram has been established.\cite{chaloupka2010, jiang2011, reuther2011, okamoto2013, schaffer2013, chaloupka2013, price2013, sela2014}
We note that bond-dependent spin interactions present in the KH model have originally been studied by Kugel and Khomskii~\cite{kugel1982} on the compass model.\cite{ khaliullin2005, nussinov2013}
However, it has turned out that the KH model cannot straightforwardly explain a zigzag-type AFM order observed in Na$_2$IrO$_3$.\cite{choi2012, ye2012}
This discrepancy has inspired further studies about more suitable effective spin models for Na$_2$IrO$_3$.
For example, further neighbor Heisenberg interactions\cite{kimchi2011, albuquerque2011, choi2012} and anisotropic interactions due to trigonal distortions\cite{bhattacharjee2012, rau2014, yamaji2014, katukuri2014, rau2014b, sizyuk2014, kimchi2014} have been introduced to the KH model to explain the zigzag order. 
In addition, a recent neutron scattering experiment has reported that magnetic order of another iridate Li$_2$IrO$_3$ is an incommensurate spiral-type order.\cite{kimchi2014,reuther2014} It is also interesting to study such an order in the KH models extended by such interactions.

Motivated by these previous studies, we examine an extended KH model including such anisotropic interactions.
We make a phase diagram of the model around Kitaev spin-liquid phase.
We find a ferromagnetic (FM) phase, a 120$^\circ$ AFM phase, two kind of incommensurate (IC) phases, and zigzag-type AFM phase next to the Kitaev spin-liquid phase.
The zigzag phase exhibits spin-spin correlation similar to a model more realistic for Na$_2$IrO$_3$.\cite{yamaji2014}

Furthermore, we investigate entanglement entropy and entanglement spectrum of the extended KH model.
We find that the lowest level of entanglement spectrum at magnetically ordered states is non-degenerate. This is clearly in contrast to the Kitaev spin-liquid state, where all of entanglement levels form pairs.
Such a degenerate structure in the Kitaev spin liquid is due to its gauge structure coming from topological nature and depends on boundary conditions.
As a result, Schmidt gap defined as the energy difference between the lowest and first excited entanglement levels changes at phase boundary between the Kitaev spin-liquid and other magnetically ordered phases.
However, we find that the Schmidt gap cannot be a good measure of phase transition between magnetically ordered phases.

This paper is organized as follows.
The extended KH model and density-matrix renormalization group (DMRG) method are introduced in Sec.~\ref{sec2}.
In Sec.~\ref{sec3}, we show a phase diagram of the model around the Kitaev spin-liquid phase, obtained by the spin-spin correlation functions and the ground state energy.
The behavior of the ground-state energy, entanglement entropy and entanglement spectrum across phase boundaries are also shown.
In addition, we discuss entanglement spectrum of the extended KH model and clarify the relations between Schmidt gap and phase transition in the model.
Finally, summary and outlook are given in Sec.~\ref{sec4}.
The entanglement spectrum of the KH model is discussed in Appendix~\ref{appendix}.

\section{Model and Method}
\label{sec2}

The Hamiltonian of an extended KH model is given by 
\begin{eqnarray}
\mathcal{ \hat H} &=& \sum_\Gamma \sum _{\langle lm \rangle \in \Gamma} \mathcal{ \hat H}_{lm} \label{simple} \qquad \\
\mathcal{\hat H}_{lm} &=& K S_l^\gamma S_m^\gamma +J \left( S_l^\alpha S_m^\alpha +S_l^\beta S_m^\beta \right) \nonumber \\
 &+& I_1 \left( S_l^\alpha S_m^\beta + S_l^\beta S_m^\alpha \right) \\
 &+&I_2 \left( S_l^\alpha S_m^\gamma + S_l ^\gamma S_m^\alpha + S_l ^\beta S_m^\gamma + S_l^\gamma S_m^\beta \right) , \nonumber
\end{eqnarray}
where $\Gamma$ represents a combination of  $(\alpha , \beta , \gamma )  =(x,y,z)$, $(z,x,y)$, and $(y,z,x)$ on the $\mathcal{J}_z$, $\mathcal{J}_y$ and $\mathcal{J}_x$ bond, respectively, and $\langle lm \rangle$ sums over all possible bonds belonging to $\Gamma$. 
We note that $I_1$ and $I_2$ terms are added to a KH model consisting of the $K$ and $J$ terms.
The $I_1$ term mainly originates from a feature of an edge-shared octahedron with total angular momentum $j=1/2$ and the $I_2$ term originates mainly from trigonal distortions present in Na$_2$IrO$_3$.
This model (\ref{simple}) has been studied by Rau and Kee\cite{rau2014b} as an effective model describing Na$_2$IrO$_3$.

We calculate the ground state of this model by using DMRG method.\cite{white1992, schollwock2005}
The DMRG calculations are carried out under periodic boundary conditions.
We take the $x$-axis direction along an armchair-edge direction and the $y$-axis direction along  a zigzag-edge direction as shown in Fig.~\ref{honey}.
Unless otherwise noted, we use a system with 6 (along $y$-axis) $\times$ 8 (along $x$-axis) sites, i.e., a 48-site system.
To perform DMRG, we construct a snake-like one-dimensional chain by combining the eight zigzag lines along $y$-axis, leading to a spin chain with long-range interactions.
We keep 1000 states in the DMRG block and performed more than 10 sweeps, resulting in a typical truncation error $5 \times 10^{-6}$ or smaller.

\section{Calculated Results and Discussions}
\label{sec3}

Putting $I_1=I_2=0$ into the extended KH model (\ref{simple}) leads to the KH model, whose phase diagram has been established.\cite{chaloupka2010, jiang2011, reuther2011, okamoto2013, schaffer2013, chaloupka2013, price2013, sela2014} 
In the phase diagram, a Kitaev spin liquid phase emerges in the range of $K/J \le -11$, when $K<0$ and $J>0$.
An interesting issue of the extended KH model concerning Na$_2$IrO$_3$ is to find a zigzag-type AFM phase around the Kitaev spin-liquid phase.\cite{rau2014b}
Since the zigzag-type AFM phase is next to the Kitaev spin-liquid phase in the parameter region of $K<0$ and $J>0$, we use these signs in the present paper. It is also interesting to investigate the $K>0$ and $J<0$ region, but it will be a future issue.

\begin{figure}[t]
\begin{center}
\includegraphics[width=18pc]{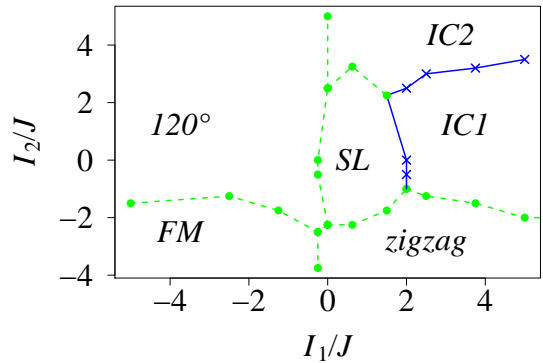}
\caption{(Color online) Phase diagram of the extended KH model (\ref{simple}). There are a ferromagnetic phase (FM), a 120$^\circ$ AFM phase (120$^\circ$), a Kitaev spin-liquid phase (SL), two incommensurate phases (IC1, IC2), and zigzag-type antiferromagnetic phase (zigzag).  The circle and X points are determined by the second derivative of energy with respect to $I_2$ and connected by lines. The boundaries denoted by blue solid lines are expected to be of first-order transition and those by green broken lines to be of continuous transition.}
\label{phase}
\end{center}
\end{figure}

\begin{figure*}[htbp]
  \begin{center}
    \begin{tabular}{r}
      \begin{minipage}{0.33\hsize}
        \begin{center}
          \includegraphics[clip, width=13pc]{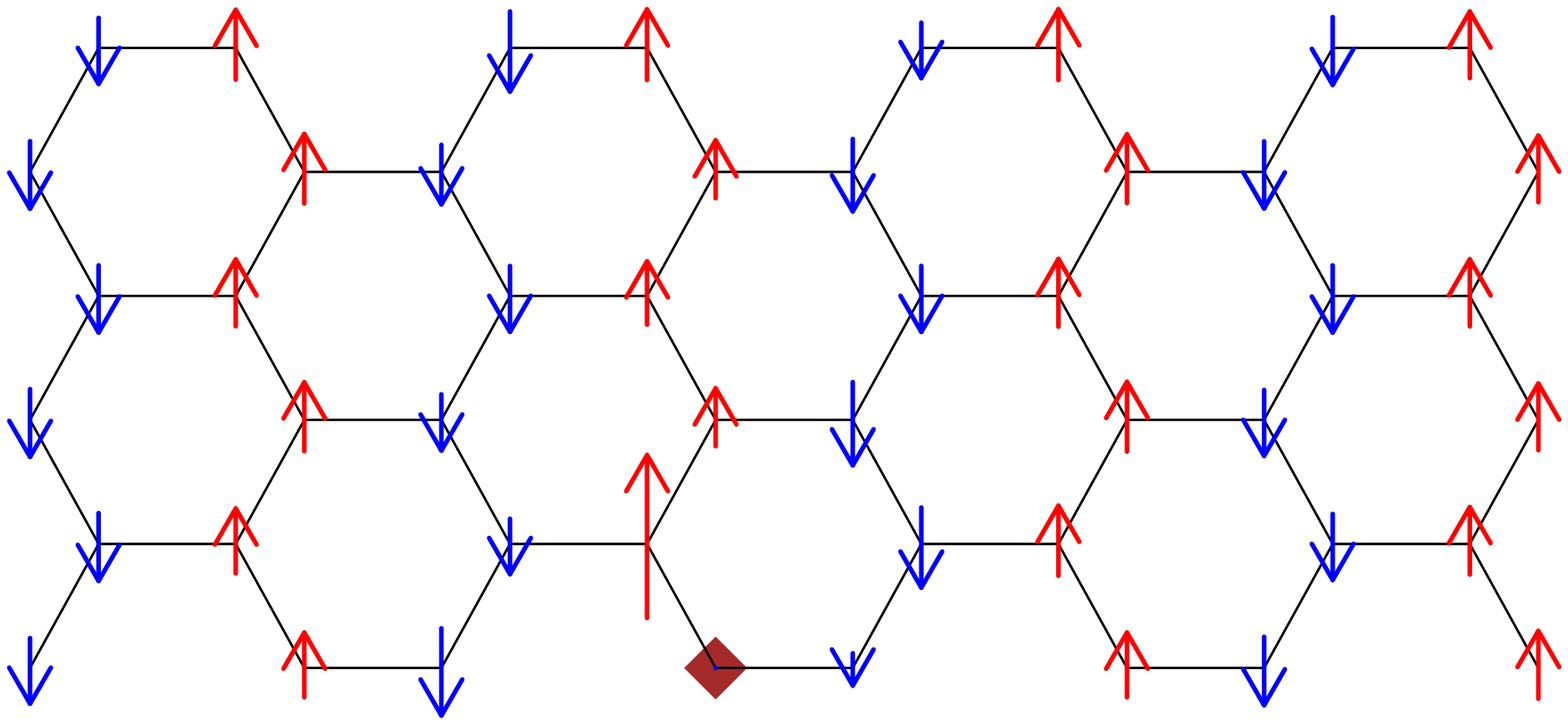}
          \hspace{1cm} (a)
        \end{center}
      \end{minipage}
      \begin{minipage}{0.33\hsize}
        \begin{center}
          \includegraphics[clip, width=13pc]{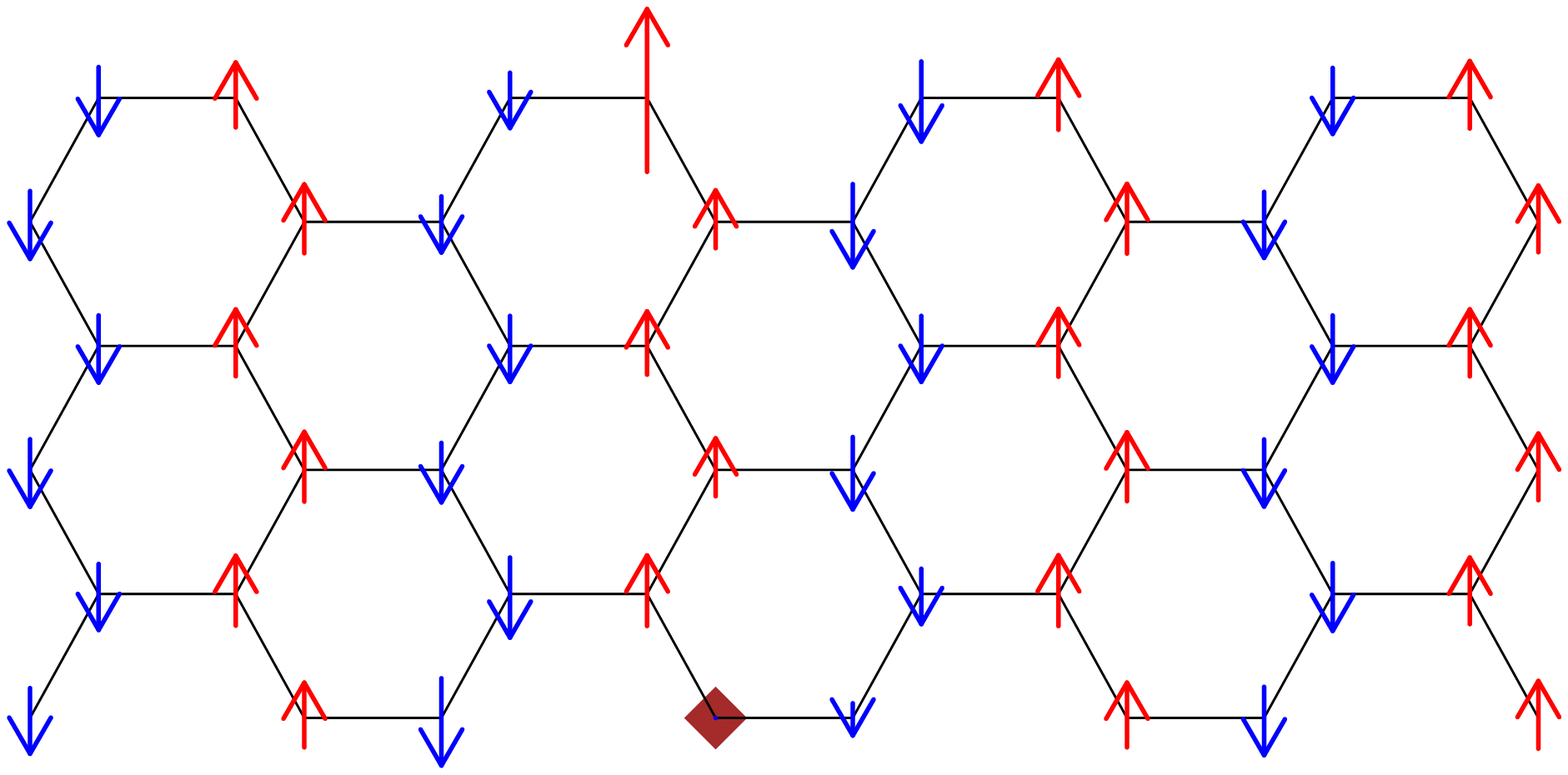}
          \hspace{1cm} (b)
        \end{center}
      \end{minipage}
      \begin{minipage}{0.33\hsize}
        \begin{center}
          \includegraphics[clip, width=13pc]{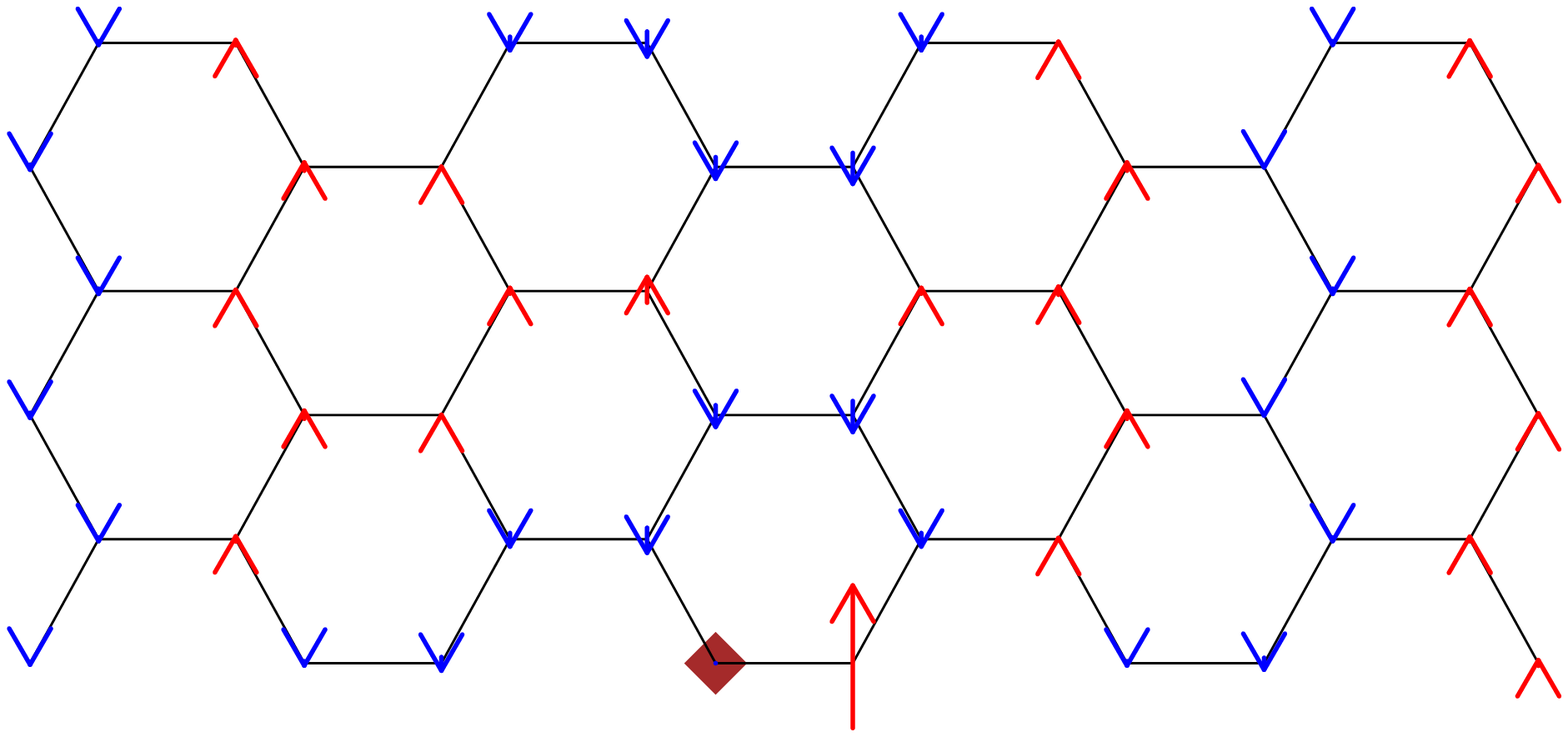}
          \hspace{1cm} (b)
        \end{center}
      \end{minipage}      
    \end{tabular}
    \caption{(Color online) (a)$\langle S_i^x S_j^x \rangle$, (b) $\langle S_i^y S_j^y \rangle $, and (c) $\langle S_i^z S_j^z \rangle $ for zigzag-type AFM phase at $I_1/J=3.8$ and $I_2/J=-3.8$.
The $i$ site is indicated by a brown rhombus point.
Upward red arrows and downward blue arrows show positive and negative values of spin-spin correlation, respectively.  
The length of the arrows represents the strength of spin-spin correlation.}
    \label{spinspin_zigzag}
  \end{center}
\end{figure*}

Fixing $K/ J =-25$, we find a zigzag-type AFM phase next to the Kitaev spin-liquid phase when $I_1/J>0$ and $I_2/J<0$ as shown in Fig.~\ref{phase}.
The presence of the zigzag state is confirmed by examining the spin-spin correlation functions for each component between sites $i$ and $j$, given by $\langle S_i^x S_j^x \rangle = \langle 0| \hat S_i^x\hat S_j^x |0\rangle$, $\langle S_i^y S_j^y \rangle = \langle 0| \hat S_i^y\hat S_j^y |0\rangle$, and $\langle S_i^z S_j^z \rangle = \langle 0| \hat S_i^z\hat S_j^z |0\rangle$, where $|0\rangle$ is the ground-state wave function.
Figure~\ref{spinspin_zigzag} shows the calculated spin-spin correlation functions for the 48-site cluster at $I_1/J=3.8$ and $I_2/J=-3.8$ in the zigzag phase.
In the figure, the $i$ site is indicated by a brown rhombus point.
Upward red arrows and downward blue arrows denote positive and negative values of spin-spin correlation, respectively.
The length of the arrows shows the absolute value of spin-spin correlation.
We find the same sign within the zigzag line along the $y$ direction in both $\langle S_i^x S_j^x \rangle$ (Fig.~\ref{spinspin_zigzag}(a)) and $\langle S_i^y S_j^y \rangle$ (Fig.~\ref{spinspin_zigzag}(b)), indicating the presence of the zigzag order.
We note that $\langle S_i^z S_j^z \rangle$ is very short-range.

It is interesting to examine whether the zigzag phase smoothly connected to that obtained by a more realistic effective spin model for Na$_2$IrO$_3$. 
Very recently, Yamaji {\it et al.} have proposed such a model based on the electronic states obtained by the first-principles calculation.\cite{yamaji2014}
By performing DMRG calculations, we have confirmed that the zigzag AFM phase in the effective model\cite{yamaji2014} exhibits spin-spin correlation similar to that shown in Fig.~\ref{spinspin_zigzag} and a similar value of the nearest neighbor spin-spin correlations.
In addition, by changing parameters continuously, we have checked that there is no phase transition between the zigzag phases of the effective model and our extended KH model.
Therefore, we can say that the zigzag phase in the effective model is smoothly connected to the zigzag phase in Fig.~\ref{phase}. 

In addition to the zigzag phase, we find various magnetic phases surrounding the Kitaev spin liquid in Fig.~\ref{phase}, which is similar to the results obtained by classical analysis and exact diagonalization calculations.\cite{yamaji2014} In the following, we discuss the details of each phase and phase boundaries.

\begin{figure}[tbp]
\includegraphics[width=0.4\textwidth]{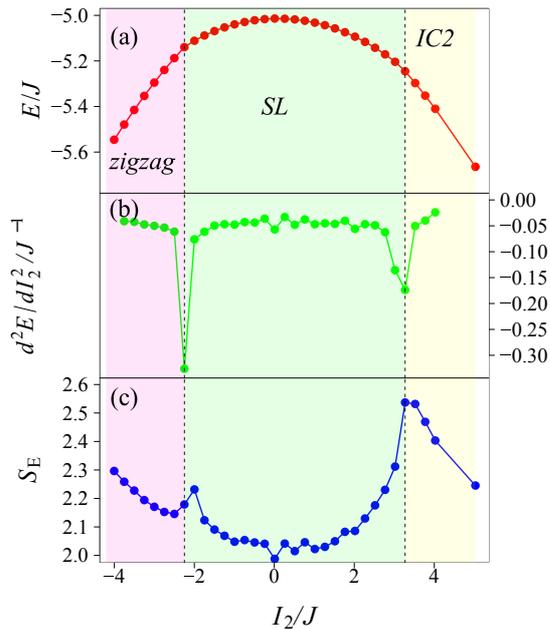}
\caption{(Color online) (a) The ground-state energy per site, $E$ (red plots), (b) second derivative of $E$ with respect to $I_2$, ${d^2 E}/{dI_2^2}$ (green plots), and (c) entanglement entropy (blue plots). $I_1 / J=0.63$. The vertical dotted lines denote phase boundary determined by the second derivative of $E$.}
\label{energy2}
\end{figure}

Firstly, we examine the case where $I_1/J=0.63$.
With increasing $I_2$, the zigzag-type AFM phase changes to an incommensurate phase denoted by IC2 through the Kitaev spin-liquid phase.
Figure~\ref{energy2}(a) shows the ground-state energy $E$ per site.
The second derivative of $E$ with respect to $I_2$ is shown in Fig.~\ref{energy2}(b). 
We can define phase transition points from the second derivative.
At $I_2 /J =-2.2$, the zigzag phase changes to the spin-liquid phase. 
The transition seems to be continuous, i.e., of second order.
However, there remains a possibility to be of weakly first order.
In order to confirm this, we need to examine the energy profile in mode detail.
This remains as a future problem.

\begin{figure*}[tbp]
  \begin{center}
    \begin{tabular}{r}
      \begin{minipage}{0.33\hsize}
        \begin{center}
          \includegraphics[clip, width=13pc]{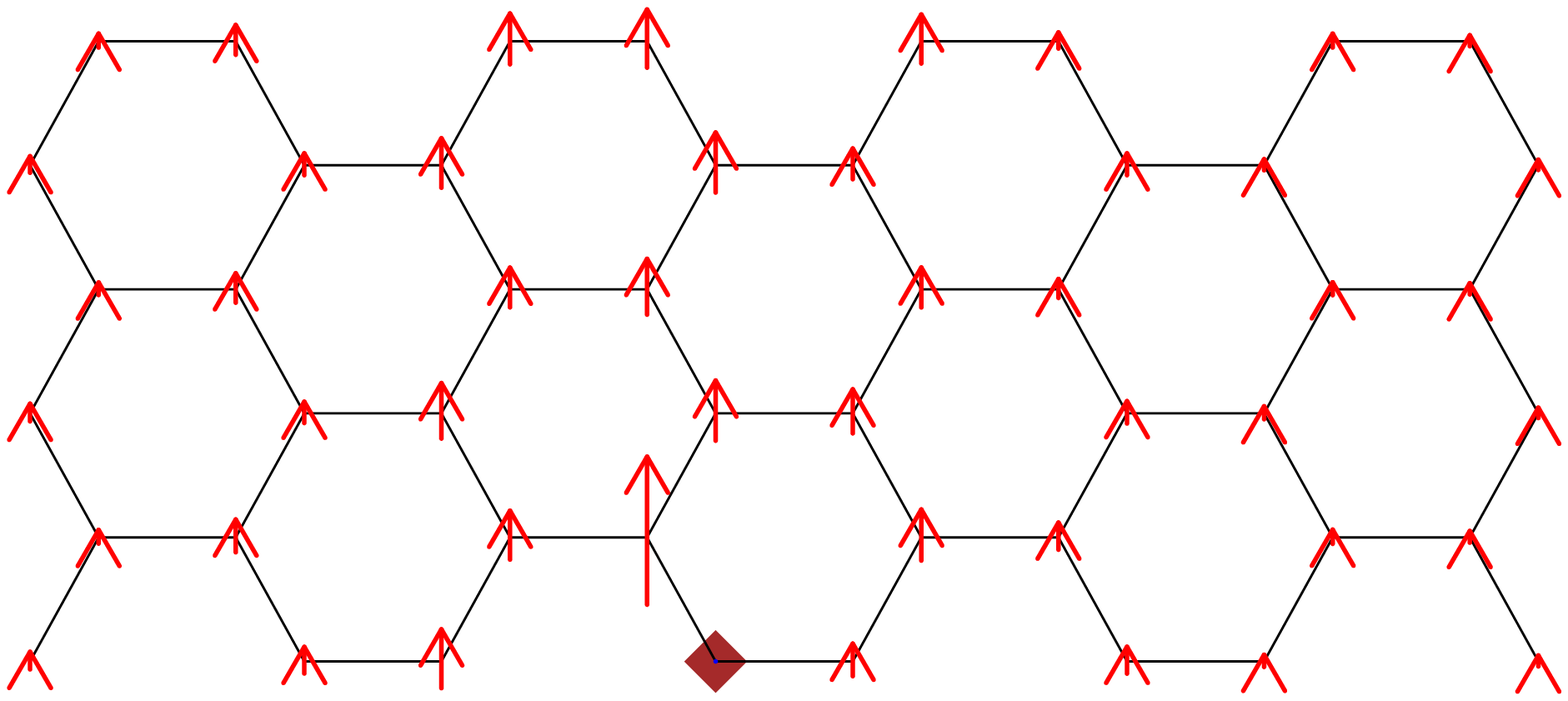}
          \hspace{1cm} (a)
        \end{center}
      \end{minipage}
      \begin{minipage}{0.33\hsize}
        \begin{center}
          \includegraphics[clip, width=13pc]{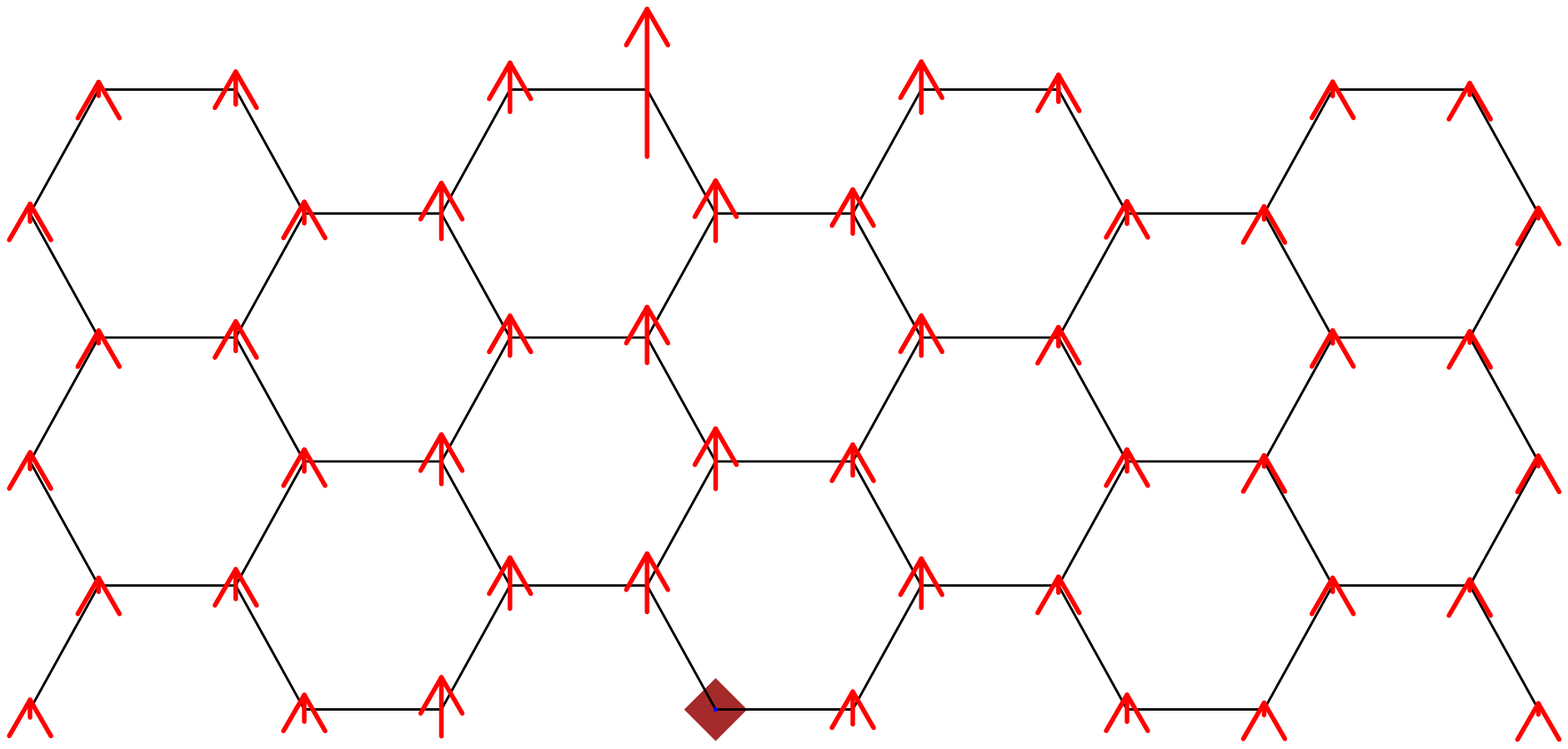}
          \hspace{1cm} (b)
        \end{center}
      \end{minipage}
      \begin{minipage}{0.33\hsize}
        \begin{center}
          \includegraphics[clip, width=13pc]{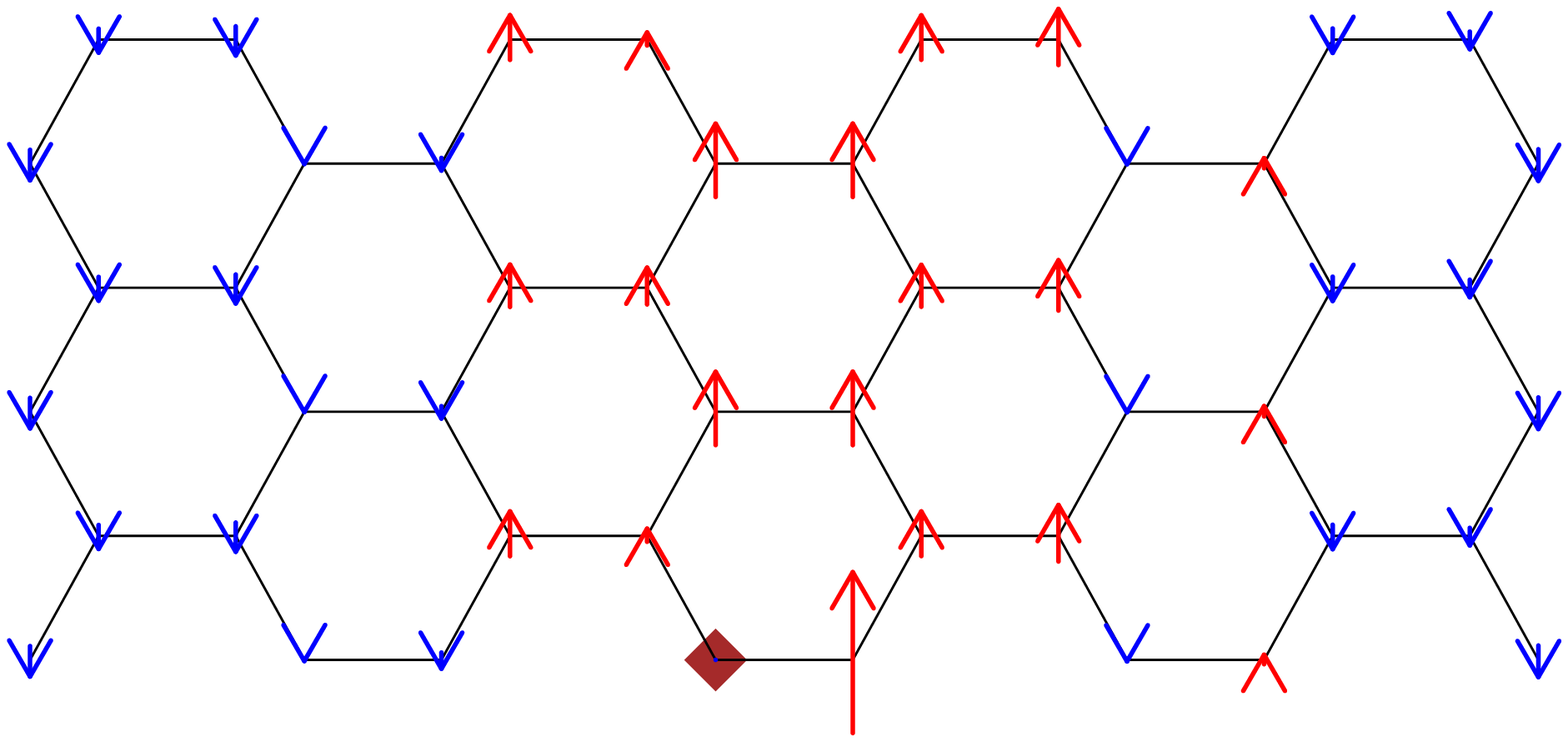}
          \hspace{1cm} (c)
        \end{center}
      \end{minipage}
    \end{tabular}
    \caption{(Color online) (a)$\langle S_i^x S_j^x \rangle$, (b) $\langle S_i^y S_j^y \rangle $, and (c) $\langle S_i^z S_j^z \rangle $ for IC2 phase at $I_1/J=2.5$ and $I_2/J=5.0$.
The $i$ site is indicated by a brown rhombus point.
Upward red arrows and downward blue arrows show positive and negative values of spin-spin correlation, respectively.  
The length of the arrows represents the strength of spin-spin correlation.}
    \label{spinspin_eff2}
  \end{center}
\end{figure*}

With further increasing $I_2$, the spin-liquid phase changes another phase at $I_2/J=3.1$.
The spin-spin correlation of the phase is shown in Figs.~\ref{spinspin_eff2}(a), \ref{spinspin_eff2}(b), and \ref{spinspin_eff2}(c). 
The correlations of $x$ and $y$ spin components show the same sign for all sites, but the $z$ component exhibits a different behavior where sign depends on distance from the $i$ site. This implies the presence of an incommensurate spin-spin correlation. We cannot clarify its propagation vector, since the system size we use is too small to determine it. We denote this phase as IC2.

Entanglement of wave function can provide useful information on quantum states.
It is measured by entanglement entropy and entanglement spectrum.\cite{li2008}
In a system composed of two subsystems A and B, a Schmidt decomposition of a many-body state $|\psi \rangle$ reads
\begin{equation}
|\psi \rangle = \sum _i p_i |\psi^i_A \rangle |\psi ^i _B \rangle
                   =\sum _i e^{-\xi_i} |\psi _A^i \rangle |\psi _B^i \rangle ,
\end{equation}
where $p_i$ is the eigenvalue of reduced density-matrix $\rho _A ={\rm Tr}_B |\psi \rangle \langle \psi | =e^{-\mathcal{H}_E}$ for subsystem A (or $\rho _B ={\rm Tr}_A |\psi \rangle \langle \psi |$ for subsystem B).
The distribution of $\xi_i$ is called entanglement spectrum, where $\xi_i$ is the eigenvalue of entanglement Hamiltonian $\mathcal{H}_E$.
Then, von Neumann entanglement entropy containing non-local topological properties can be written as 
\begin{equation}
S_\mathrm{E} =-\sum _i p_i \ln p_i = \sum _i \xi _i e^{-\xi _i}.
\end{equation}
We take the A subsystem be half of the whole system throughout this paper.
When we consider a system with toroidal geometry coming from periodic boundary conditions, we cut the whole system twice.
In cylindrical geometry we divide a system into two subsystems.

In Fig.~\ref{energy2}(c), $S_\mathrm{E}$ for $I_1/J=0.63$ is shown.
$S_\mathrm{E}$ shows a peak structure near phase boundary but the peak position is not exactly at the boundary.
This is clearly seen at $I_2/J\simeq-2$, where there is the boundary between the zigzag and spin-liquid phases.
There have been many studies about the relationship between entanglement entropy and phase transition in one-dimensional systems,\cite{holzhey1994, calabrese2004} showing a diverging behavior in $S_\mathrm{E}$ at phase transition points.
However, such relationship has not yet been established in two-dimensional systems.
Therefore, we need to make clear whether the relationship is applicable for our system or not.
For this purpose, entanglement spectrum may be helpful for understanding the behavior of entanglement entropy at phase boundary.

\begin{figure}[t]
\begin{center}
\includegraphics[width=18pc]{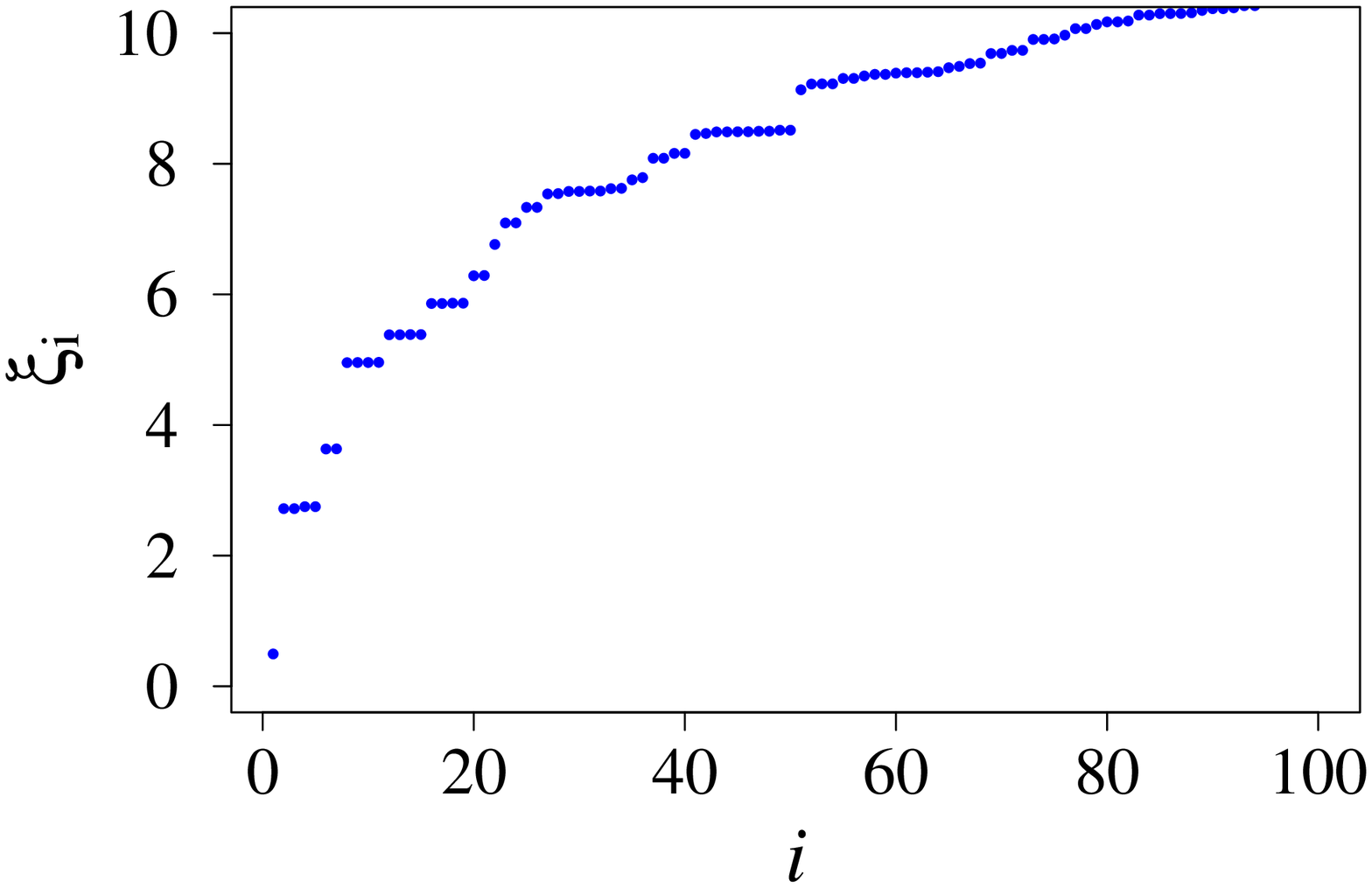}
\caption{Entanglement spectrum for a zigzag-type AFM ordered ground state at $I_1/J=3.8$ and $I_2/J=-3.8$.}
\label{es_zigzag}
\end{center}
\end{figure}

Before discussing entanglement spectrum near phase boundary, we show the spectrum for a zigzag-type ordered state ($I_1/J=3.8$ and $I_2/J=-3.8$) in Fig.~\ref{es_zigzag}, where entanglement levels are plotted from the smallest value starting from $i=1$. 
The lowest level of the entanglement spectrum $\xi_1$ is non-degenerate and separated from $\xi_2$. 
In the following, we call the level separation $\xi_2-\xi_1$ the Schmidt gap.
We note that the non-degenerate $\xi_1$ is clearly in contrast to the Kitaev spin-liquid state, where all of entanglement levels form pairs (See Appendix~\ref{appendix}).

\begin{figure}[tbp]
\includegraphics[width=18pc]{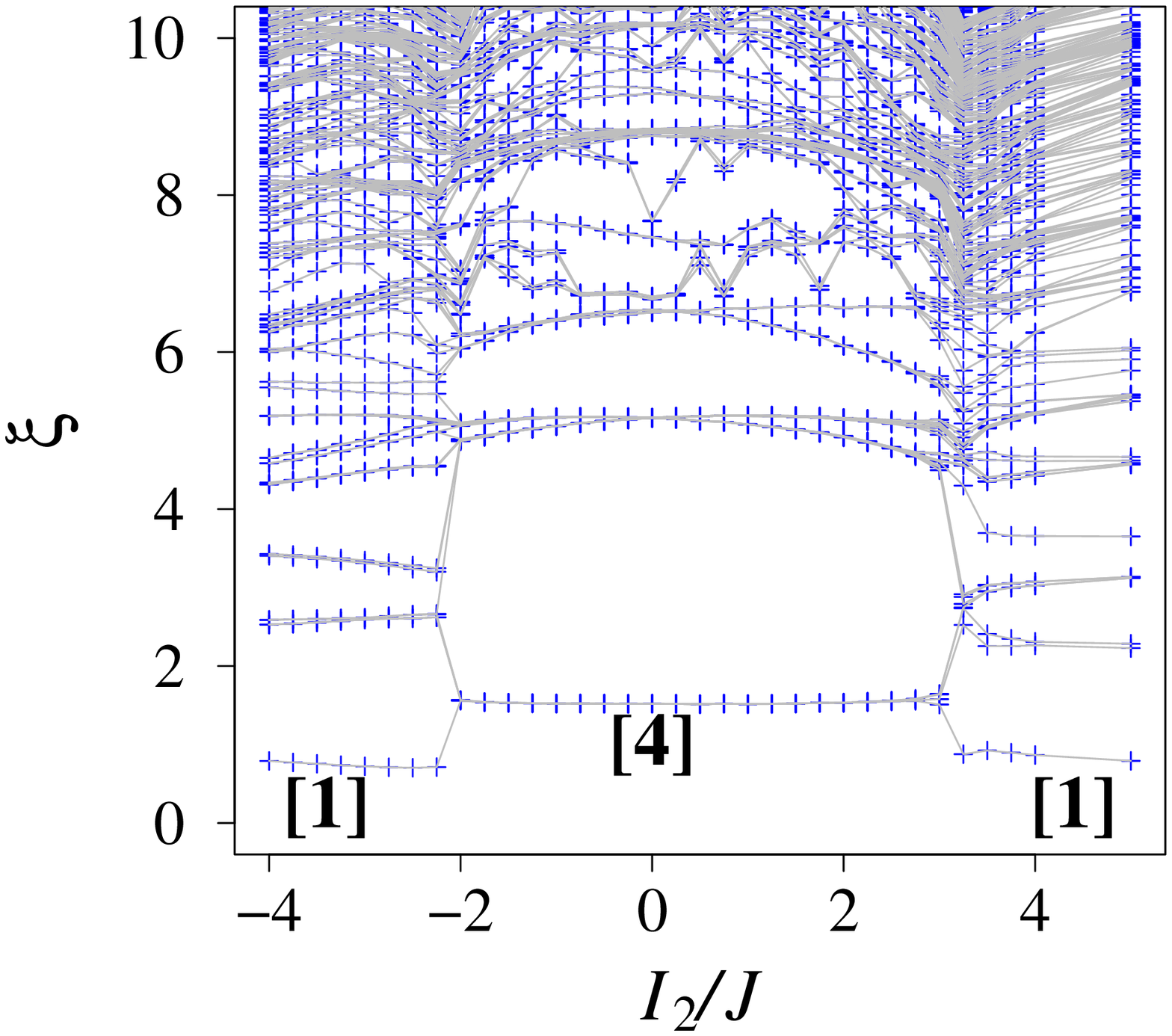}
\caption{(Color Online) Entanglement spectrum for the extended KH model (\ref{simple}). $I_1 / J=0.63$. Blue crosses represent entanglement levels and gray lines connect the spectrum belonging to the same entanglement levels. $[n]$ denotes $n$-fold degeneracy of the lowest entanglement level in each phase.}
 \label{es_eff2}
\end{figure}

The spectral distribution of the entanglement spectrum changes with changing parameters.
Figure~\ref{es_eff2} shows entanglement spectrum for the $I_1=0.63$ case corresponding to Fig.~\ref{energy2}.
We find that the Schmidt gap changes from zero to finite at $I_2 /J =-2.2$ ($I_2 / J = 3.2$) with decreasing (increasing) $I_2$ from the spin-liquid phase.
These $I_2$ values are consistent with the transition points obtained by the second derivative of $E$.
Comparing $I_2 /J =-2.2$ with the peak position of $S_\mathrm{E}$ ($I_2 /J =-2.0$), we may judge that the Schmidt gap is more appropriate than the entanglement entropy for the determination of the phase boundary in two-dimensional systems.
Of course, more studies on different systems are necessary to confirm this statement.
We also note that there is a different case where the Schmidt gap itself cannot be a measure of phase transition, as will be discussed below.

Here, we comment on the degeneracy of $\xi_i$ in the Kitaev spin-liquid phase, which is located at the middle region of Fig.~\ref{es_eff2}. [$n$] in this figure shows the number of degeneracy of the lowest entanglement level and [$4$] in the Kitaev spin-liquid phase denotes 4-fold degeneracy.
As discussed in Appendix~\ref{appendix}, this is due to the gauge structure of the Kitaev spin liquid.
We consider that the degeneracy is one of the ``fingerprint" of the Kitaev spin liquid.
Such gauge structure also appears in topological entanglement entropy,~\cite{yao2010} and thus the degeneracy of entanglement spectrum is useful for characterizing the nature of spin-liquid phase.
We discuss the entanglement spectrum of the Kitaev spin liquid in more detail in Appendix~\ref{appendix}.

\begin{figure}[tbp]
\includegraphics[width=0.4\textwidth]{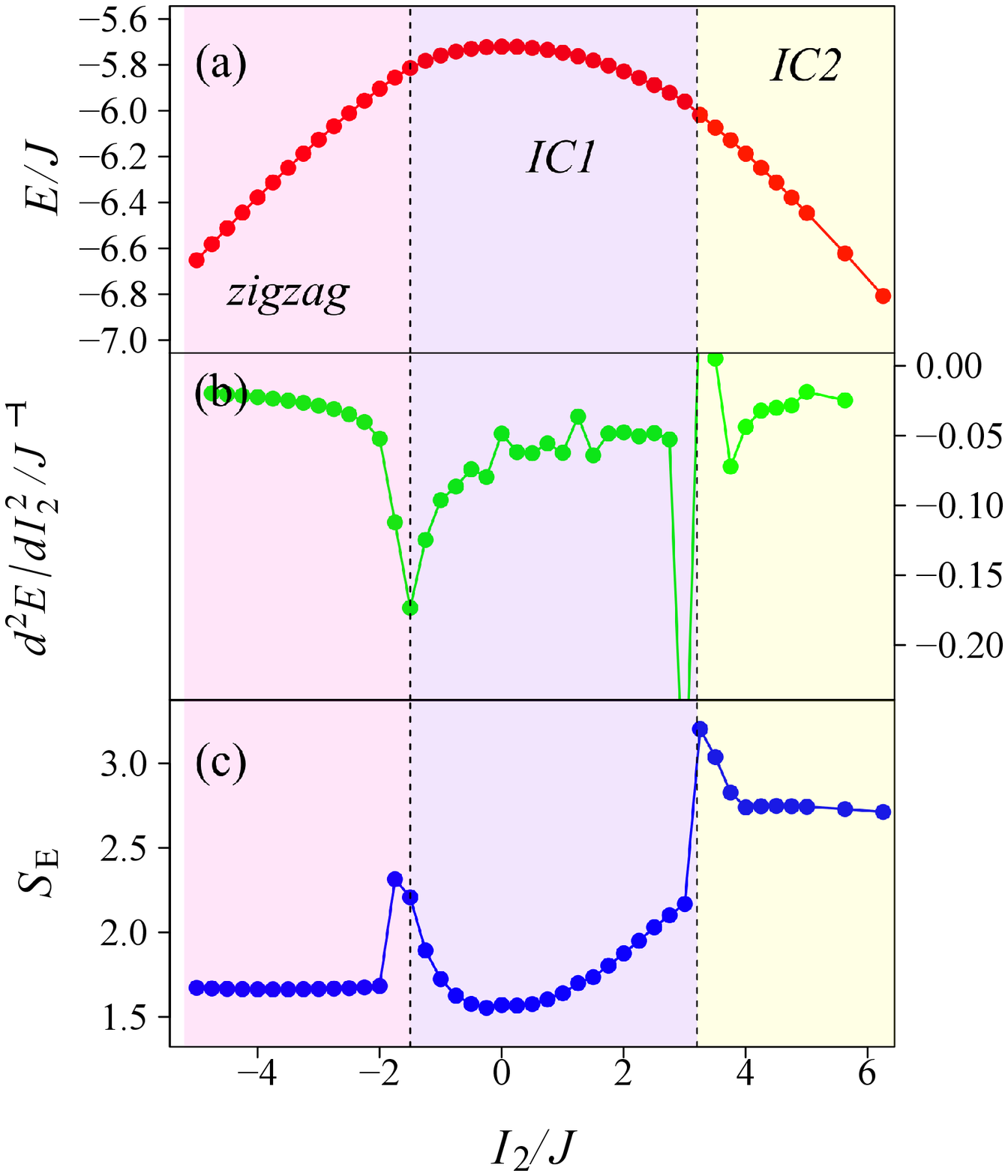}
\caption{(Color online) Same as Fig.~\ref{energy2}, but $I_1 / J=3.8$.}
\label{energy3}
\end{figure}

\begin{figure*}[tbp]
  \begin{center}
    \begin{tabular}{r}
      \begin{minipage}{0.33\hsize}
        \begin{center}
          \includegraphics[clip, width=13pc]{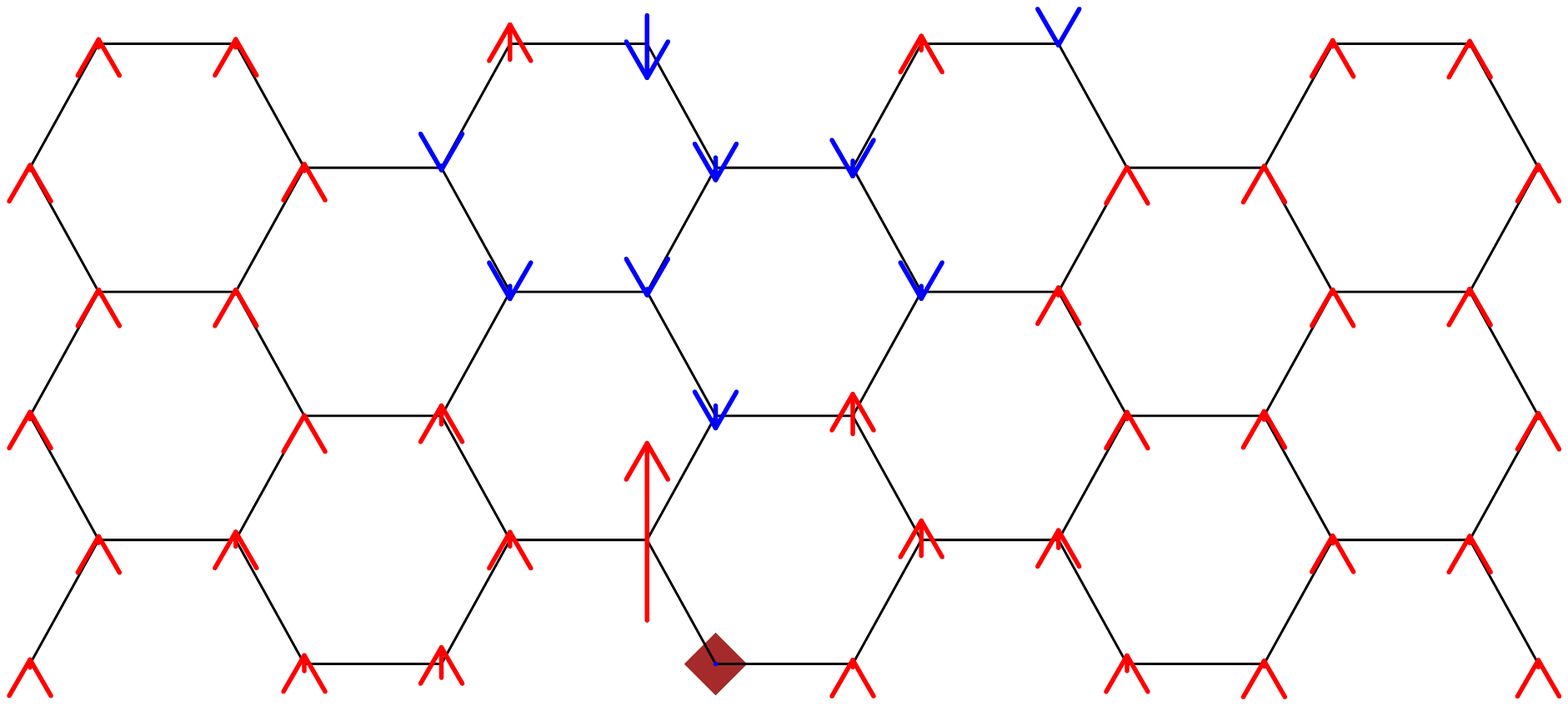}
          \hspace{1cm} (a)
        \end{center}
      \end{minipage}
      \begin{minipage}{0.33\hsize}
        \begin{center}
          \includegraphics[clip, width=13pc]{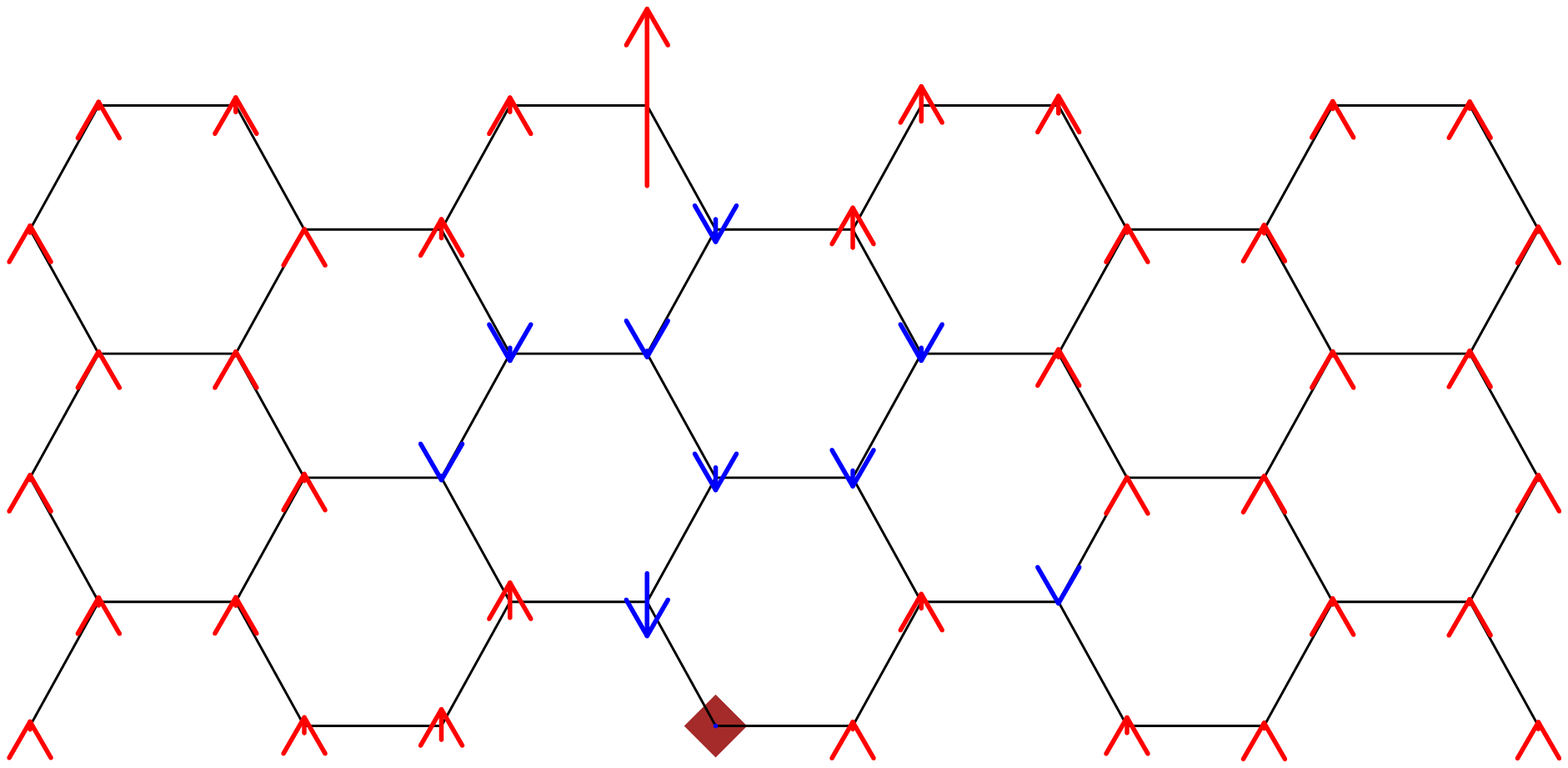}
          \hspace{1cm} (b)
        \end{center}
      \end{minipage}
      \begin{minipage}{0.33\hsize}
        \begin{center}
          \includegraphics[clip, width=13pc]{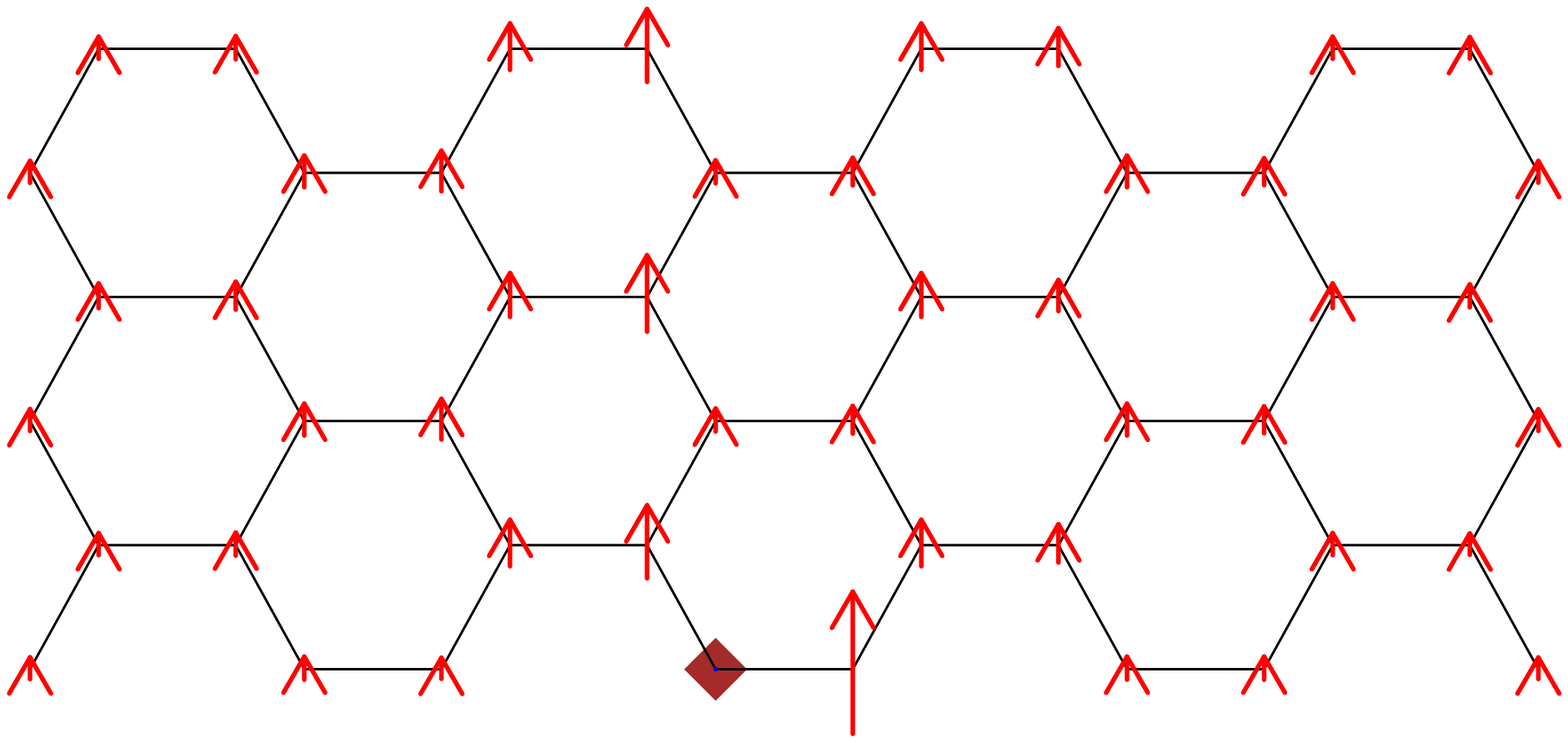}
          \hspace{1cm} (c)
        \end{center}
      \end{minipage}
    \end{tabular}
    \caption{(Color online) Same as Fig.~\ref{spinspin_eff2}, but for IC1 phase at $I_1/J=2.5$ and $I_2/J=2.5$.}
    \label{spinspin_eff1}
  \end{center}
\end{figure*}

Returning to the phase diagram in Fig.~\ref{phase}, we next examine the case where $I_1/J=3.8$.
Figures~\ref{energy3}(a) and \ref{energy3}(b) show $E$ and $d^2E/dI_2^2$, respectively.
With increasing $I_2$, the zigzag-type AFM phase changes to the IC2 phase through a new phase denoted by IC1.
The second derivative of $E$ indicates that the phase transition between the zigzag and IC1 phases at $I_2/J=-1.5$ is of continuous and that between the IC1 and IC2 phases at $I_2/J=3.2$ is of first order.
Spin-spin correlation functions in the IC1 phase are shown in Figs.~\ref{spinspin_eff1}(a), \ref{spinspin_eff1}(b), and \ref{spinspin_eff1}(c) for $\langle S_i^x S_j^x \rangle$, $\langle S_i^y S_j^y \rangle$, and $\langle S_i^z S_j^z \rangle$, respectively.
$\langle S_i^x S_j^x \rangle$ and $\langle S_i^y S_j^y \rangle$ indicate non-commensurate spin arrangement, though $\langle S_i^z S_j^z \rangle$ shows a FM correlation.
This pattern of the spin-spin correlation is different from that in IC2 shown in Fig.~\ref{spinspin_eff2}.
Therefore, we denote this phase as IC1.

\begin{figure}[tbp]
\includegraphics[width=18pc]{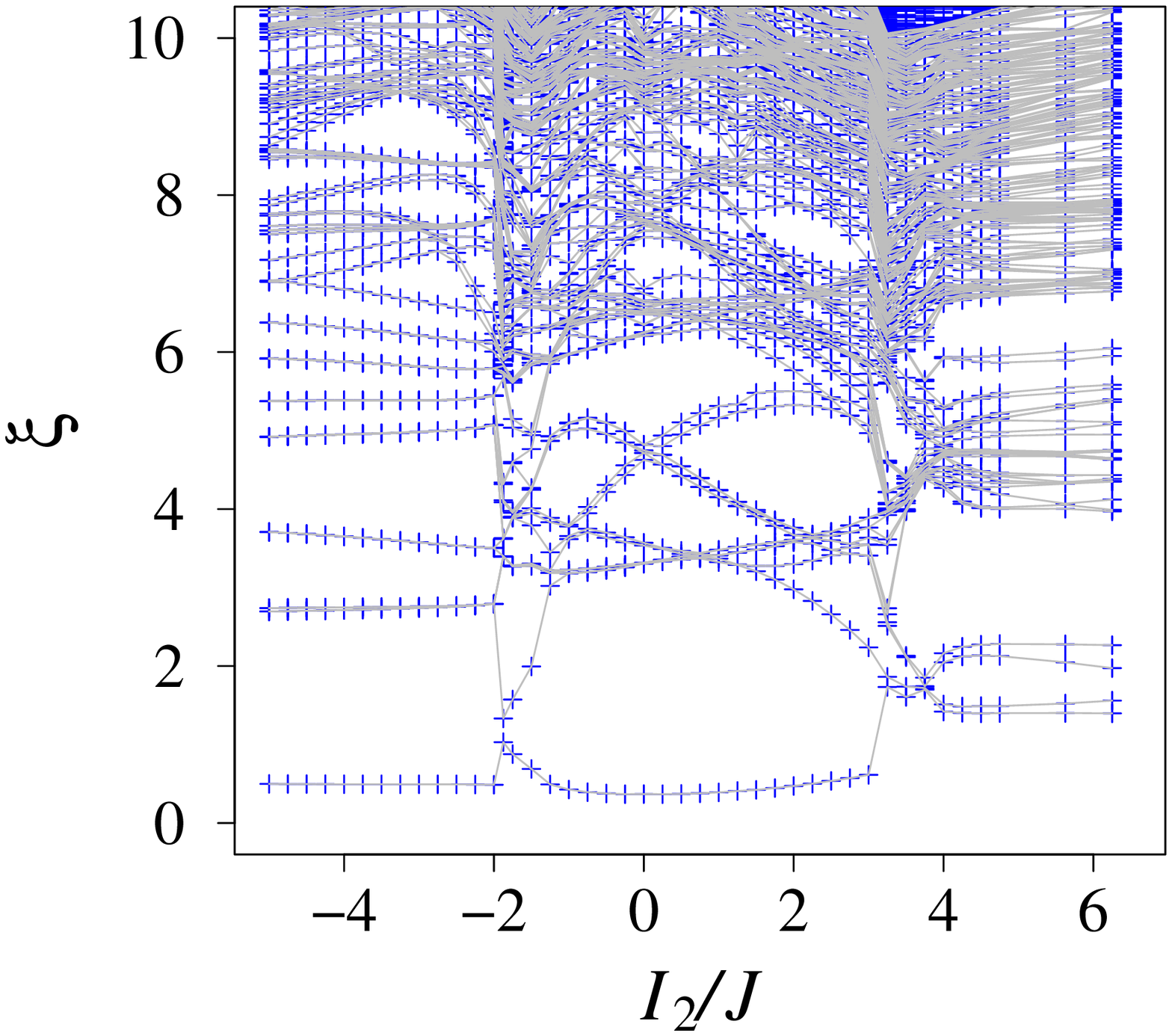}
\caption{(Color Online) Same as Fig.~\ref{es_eff2}, but $I_1 / J=3.8$.}
 \label{es_eff3}
\end{figure}

The sudden change of $S_\mathrm{E}$ at $I_2/J=3.2$ in Fig.~\ref{energy3}(c) is consistent with the first-order transition.
The entanglement spectrum and the Schmidt gap also show a change at the same value as shown in Fig.~\ref{es_eff3}.
On the other hand, the phase boundary at $I_2/J=-1.5$ disagrees with the peak position of $S_\mathrm{E}$ and also disagrees with the change of the Schmidt gap.
Such a disagreement is different from the case of the boundary between the zigzag and spin-liquid phases discussed above.

\begin{figure}[tbp]
\includegraphics[width=0.4\textwidth]{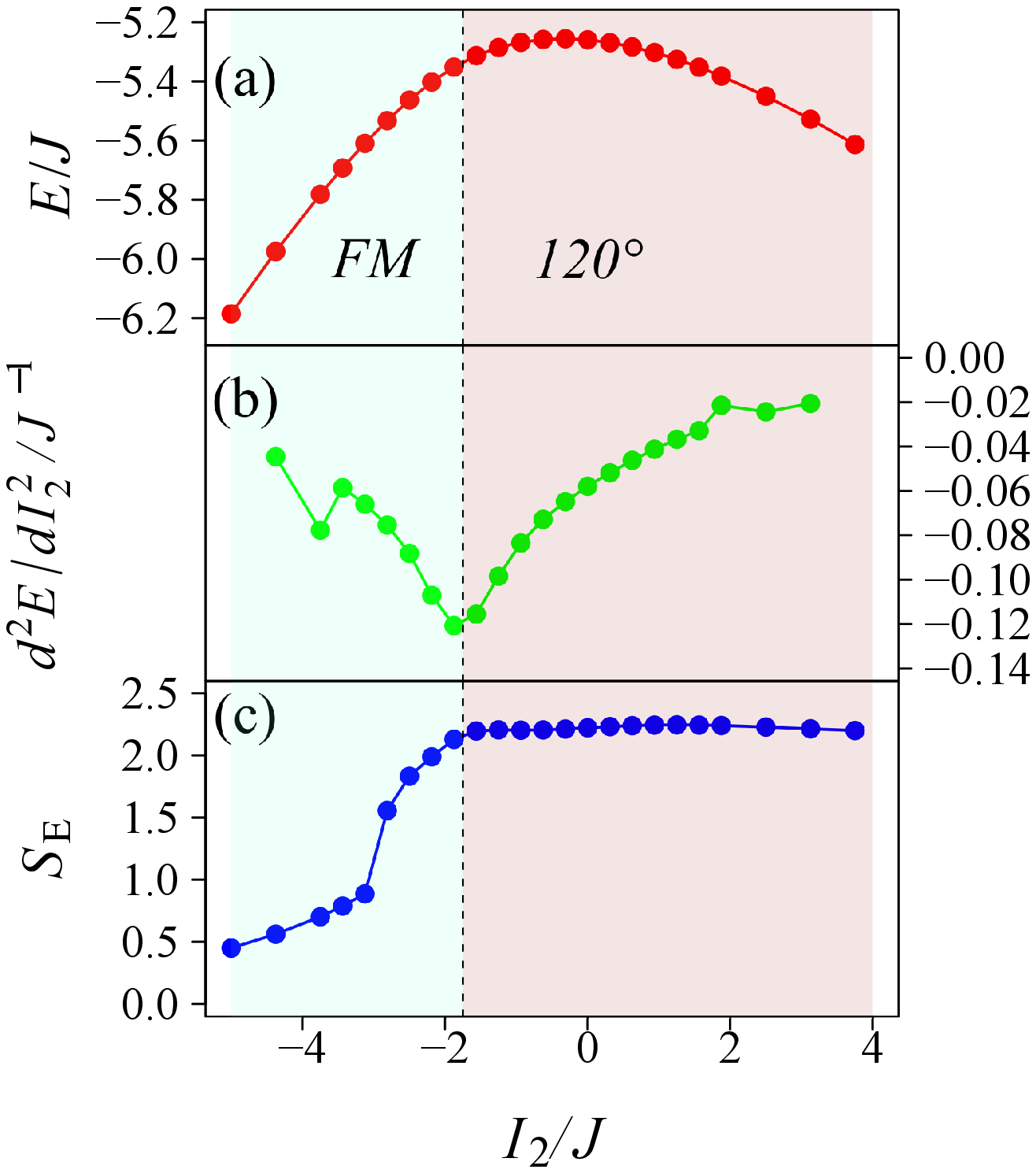}
\caption{(Color online) Same as Fig.~\ref{energy2}, but $I_1 / J=-1.3$.}
\label{energy1}
\end{figure}

Thirdly, let us examine the case of $I_1/J=-1.3$.
Figures~\ref{energy1}(a), \ref{energy1}(b), and \ref{energy1}(c) show $E$, $d^2E/dI_2^2$, and $S_\mathrm{E}$, respectively.
With increasing $I_2$, phase changes from a FM phase to a 120$^\circ$ phase at $I_2/J=-1.8$ with continuous transition.
This 120$^\circ$ phase has the same spin configuration as presented by Rau and Kee.\cite{rau2014, rau2014b}
Entanglement entropy smoothly changes at the phase boundary, in contrast to other cases where a peak structure appears.

\begin{figure}[tbp]
\includegraphics[width=18pc]{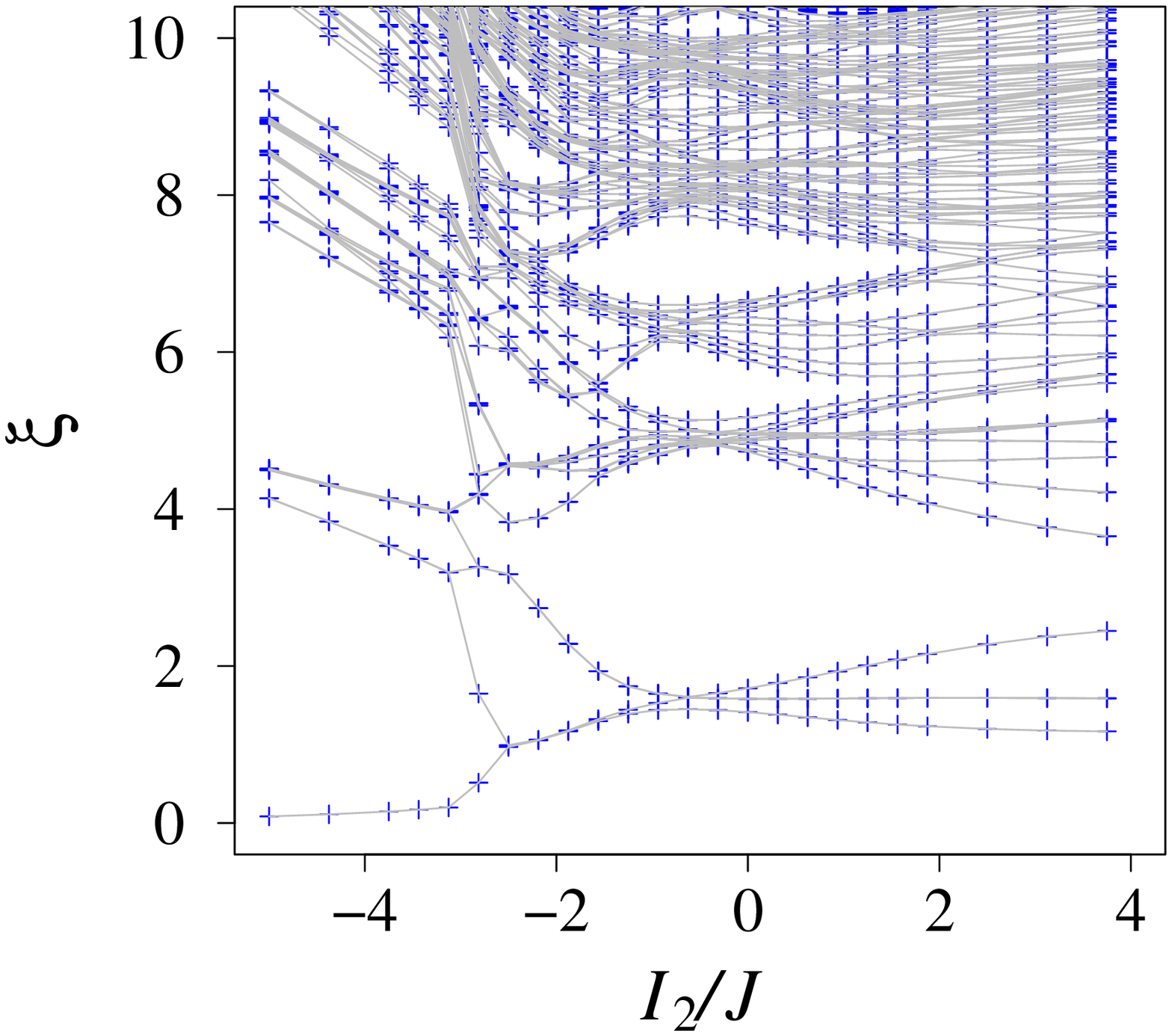}
\caption{(Color Online) Same as Fig.~\ref{es_eff2}, but $I_1 / J=-1.3$.}
 \label{es_eff1}
\end{figure}

Figure~\ref{es_eff1} shows entanglement spectrum as a function of $I_2$.
We find that the Schmidt gap changes from zero to finite at $I_2 / J = -2.8$, but there is no qualitative change at the phase boundary $I_2 / J = -1.8$.
This means that the Schmidt gap is not a good measure of the phase transition in this case where the FM phase changes to the 120$^\circ$ AFM phase.
Recently it has been shown that low-energy entanglement spectrum can exhibit singular changes, even when the physical system remains in the same phase,\cite{chandran2014} suggesting less universal information about quantum phases in the low-energy entanglement spectrum.

Therefore, we can say that our case would be such an example in two-dimensional systems, where the Schmidt gap cannot characterize phase transition points.

\section{Summary and Outlook}
\label{sec4}
We have studied the extended KH model (\ref{simple}) by using DMRG and constructed a phase diagram around the Kitaev spin-liquid phase.
We have found a FM phase, a 120$^\circ$ phase, two kinds of incommensurate phases (IC1 and IC2), and a zigzag-type AFM phase next to the Kitaev spin-liquid phase.
The zigzag phase exhibits spin-spin correlation similar to a more realistic model for Na$_2$IrO$_3$.\cite{yamaji2014}
We define phase boundaries by using the second derivative of energy.
At the boundaries, entanglement entropy does not necessarily show an anomalous behavior.
This means that the entanglement entropy is not a good measure for determining phase boundary in the extended KH model.

Examining entanglement spectrum, we have found that the lowest entanglement level in magnetically ordered states is non-degenerate. This is in contrast to that of the Kitaev spin-liquid phase, where all of entanglement levels form pairs.
We note that the degeneracy in Kitaev spin liquid is due to gauge structure and the number of its degeneracy depends on boundary condition reflecting topological nature of the Kitaev spin-liquid as discussed in Appendix~\ref{appendix}.
Therefore, phase boundaries between the Kitaev spin-liquid and the magnetically ordered phases is determined by examining entanglement spectrum.
In this case, the Schmidt gap, defined as the difference between the lowest and first-excited entanglement levels, is a useful quantity to determine the boundary.

However, as far as phase transitions between magnetically ordered phases are concerned, we have found that the Schmidt gap is not necessarily a measure of phase transition.
For example, the Schmidt gap cannot characterize phase transition between the FM and the $120^\circ$ AFM phases, between zigzag-type AFM and IC1 phases, and between the IC1 and IC2 phases.

In one-dimensional quantum many-body systems, the Schmidt gap is known to be a novel quantity for identifying and characterizing various phases and phase transitions.
In two-dimensional systems, however, the meaning of the Schmidt gap has not yet clarified as far as we know.
Therefore, we consider that the present work will provide a starting point for the study of the relation between entanglement spectrum and quantum state in two dimensions.
In fact, our present study of entanglement spectrum is closely related to other studies attempting unbiasedly to detect order parameters and/or dominant correlations using reduced density-matrices.\cite{furukawa2006,henley2014}
We believe that we are able to extract much more information from the structure of entanglement and to identify and characterize various orders more efficiently, once we understand the nature of entanglement in many-body interacting systems.

\begin{acknowledgments}
We acknowledge Y. Yamaji, S. Morita , M. Kurita, M. Imada, T. Okubo, N. Kawashima, K. Totsuka, and S. Yunoki for useful and stimulating discussions. 
This work is financially supported by MEXT HPCI Strategic Programs for Innovative Research (SPIRE) (hp130007) and Computational Materials Science Initiative (CMSI).
Numerical calculation was partly carried out at the K computer, the RIKEN Advanced Institute for Computational Science, and the Supercomputer Center, Institute for Solid State Physics, University of Tokyo.
This work was also supported by Grant-in-Aid for Scientific Research (No. 26287079 and No. 22740225) from MEXT, Japan.

\end{acknowledgments}
 
\appendix

\section{Entanglement Spectrum}
\label{appendix}

Li and Haldane proposed entanglement spectrum that contains the full set of eigenvalues of density-matrix.\cite{li2008}
Writing the eigenvalues of density-matrix as e$^{-\xi _i}$, where $\xi _i$ is an entanglement level, they have shown that the low-level entanglement spectrum for Laughlin, Moore-Read and Read-Rezayi states exhibit a universal structure related to associated conformal field theory.
The universal structure is separated from a nonuniversal high-level spectrum by entanglement gap that is finite in thermodynamic limit.
This gap itself is proposed to be a ``fingerprint" of the topological order.
Since the proposal, entanglement spectrum has been studied in various systems including fractional quantum Hall systems,\cite{li2008, regnault2009, thomale2010may, lauchli2010} topological insulators,\cite{turner2010, fidkowski2010} spin chain,\cite{thomale2010sep} and Kitaev honeycomb lattice model.\cite{yao2010}
Furthermore, it has been realized that the scaling of the Schmidt gap defined by the difference between the two largest eigenvalues of the reduced density matrix is useful for detecting critical points through the studies for one-dimensional Kugel-Khomskii model,\cite{lundgren2012} spin chains,\cite{lepori2013, giampaolo2013} and two-dimensional quantum Ising model.\cite{james2013}
Entanglement spectrum is thus now accepted to be a quantity characterizing not only various phases but also phase transitions.
However, it has recently been pointed out that the low-energy entanglement spectrum does not necessarily provide universal information about quantum phases.\cite{chandran2014}
Therefore, it is interesting to examine entanglement spectrum of the KH model whose ground state is well known.\cite{chaloupka2010}

\subsection{Kitaev-Heisenberg model}
\label{A1}
In this section, we defined the KH model as
\begin{equation}
\mathcal{H}=\sum _{\langle i,j \rangle} \left[-2\alpha S_i^\gamma S_j^\gamma + (1-\alpha )\bvec{S}_i \cdot \bvec{S}_j \right],
\label{KHmodel}
\end{equation}
where $\alpha$ is related to $K$ and $J$ in Eq.~(\ref{simple}) as $J=1-\alpha$ and $K=1-3\alpha$.
The ground state at $\alpha=0$ and $\alpha=1$ is the N\'eel and Kitaev spin-liquid state, respectively.
In between, there is a strpye-type AFM state.

\begin{figure}[t]
\begin{center}
\includegraphics[width=18pc]{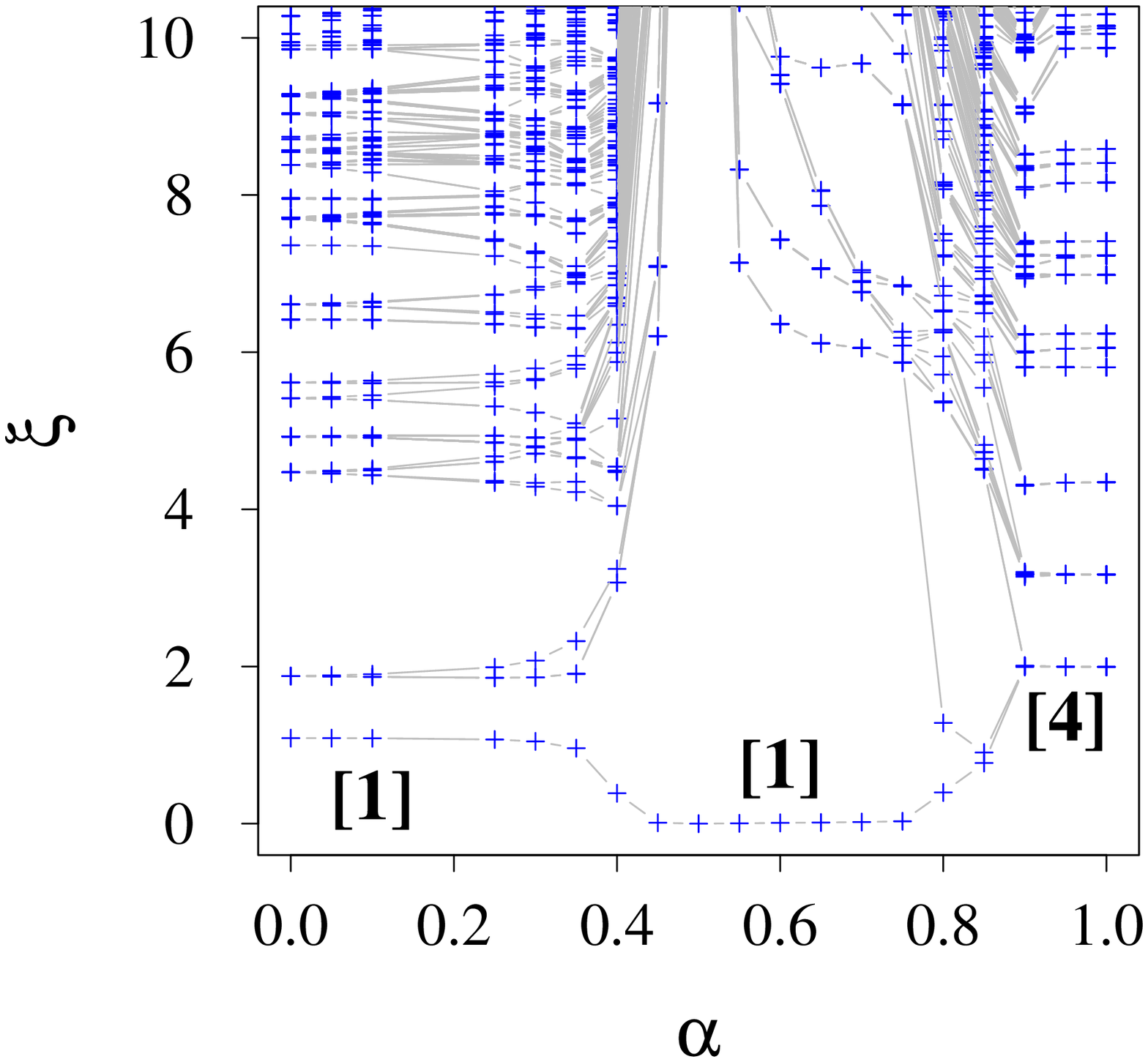}
\caption{(Color Online) Entanglement spectrum for the KH model (\ref{KHmodel}). Blue crosses represent entanglement levels and gray lines connect the spectrum belonging to the same entanglement levels. $[n]$ in (b) denotes $n$-fold degeneracy of the lowest entanglement level in each phase.}
\label{es_kh}
\end{center}
\end{figure}

Figure~\ref{es_kh} shows entanglement spectrum of the $6\times 8$-site KH model with periodic boundary conditions as a function of $\alpha$. Hereafter, we call periodic boundary condition toroidal boundary condition.
We find that level structure changes at $\alpha \simeq 0.4$ and $\alpha \simeq 0.86$.
These values are consistent with phase transition points determined by the second derivative of energy with respect to $\alpha$.

We find that the Kitaev spin-liquid phase exhibits 4-fold degeneracy in the ground state while the N\'eel and stripy phases show non-degenerate lowest energy level. 
The degeneracy of the spin-liquid phase comes from its gauge structure as will be discussed in Appendix~\ref{A2}.

The Schmidt gap increases drastically at $\alpha \simeq 0.4$ with increasing $\alpha$. 
This indicates that a phase transition occurs there. 
At exactly solvable point $\alpha=0.5$, the gap diverges, since the ground state can be written by a single product state.
With further increasing $\alpha$, the Schmidt gap closes between $\alpha=0.85$ and 9.0. This is again consistent with the position of phase boundary.
Of course, in order to determine phase boundary precisely, it is important to study finite-size scaling of Schmidt gap.

\subsection{Kitaev Spin-Liquid State}
\label{A2}
In this section, we discuss the dependence of entanglement spectrum in the Kitaev spin-liquid state on system size and boundary condition.

\begin{figure}[!t]
\begin{center}
\includegraphics[width=18pc]{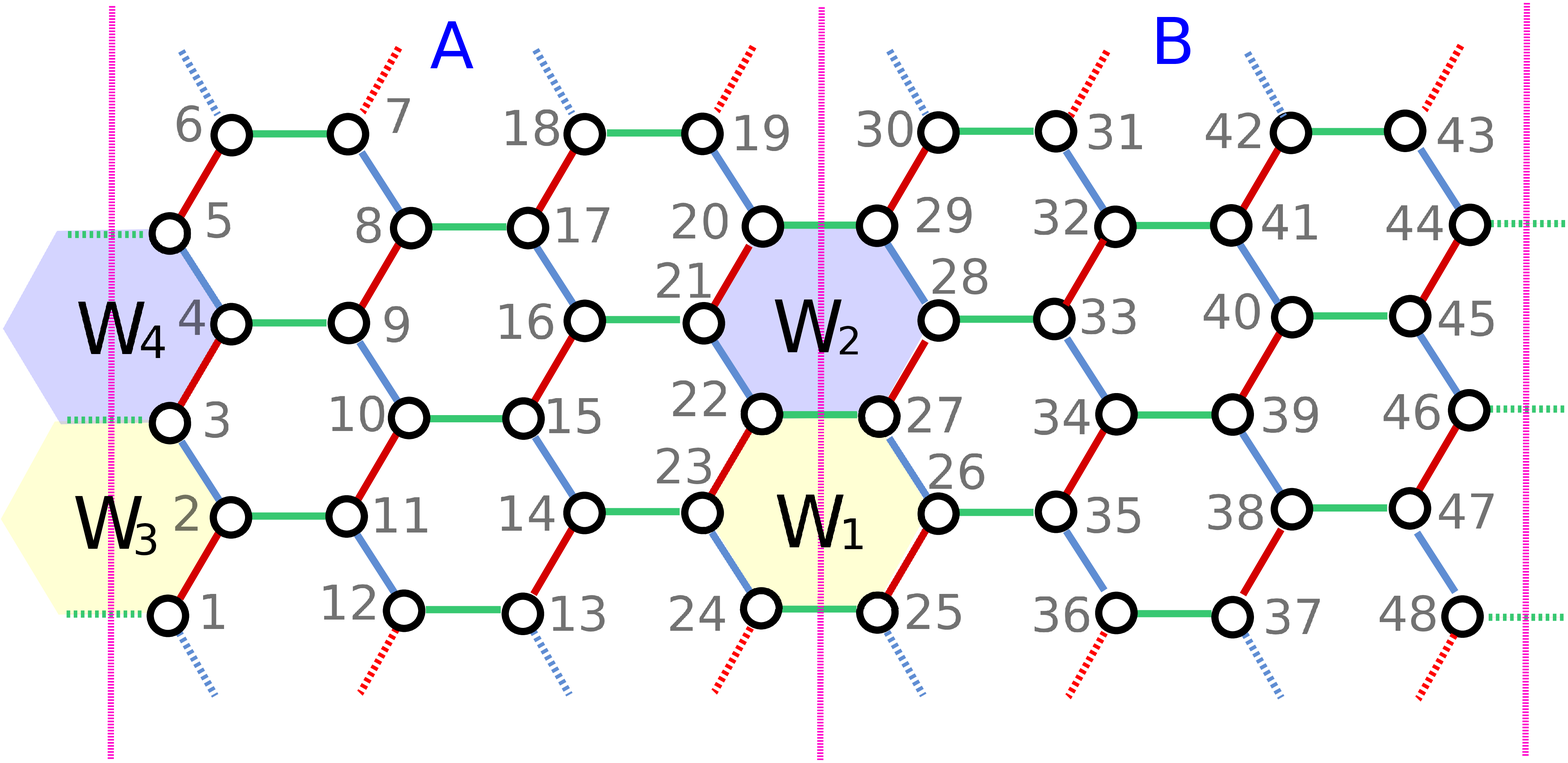}
\caption{(Color online) Cluster configuration of $6\times 8$ sites.
The numbers label sites on honeycomb lattice. Vertical dashed lines denote cutting position when we divide the whole system into two subsystems, A and B. For cylindrical boundary condition the system is divided only once at the middle vertical line, while for toroidal boundary condition the system is cut twice at the middle vertical line and the right or left vertical line. $W_1$ and $W_2$ show the Wilson loops defined on hexagon on honeycomb lattice, which cross the middle vertical line. $W_3$ and $W_4$ show the loops that cross the right or left vertical line.}
\label{wilson}
\end{center}
\end{figure}

First of all, we consider the degeneracy of entanglement spectrum for a 6$\times L_x$ ($L_x\rightarrow\infty$) system by counting the Wilson loops that are cut when the whole system is divided into two subsystems.\cite{kitaev2006, fradkin, morita}
In our cluster configuration, it is inevitable to have two Wilson loops, for example, $W_1$ and $W_2$ defined on two neighboring hexagons as shown in Fig.~\ref{wilson}.
The two loops induce 2-fold degeneracy.
The number of degeneracy increases as the number of Wilson loop defined on honeycomb lattice increases.
We can define more Wilson loops in toroidal boundary condition than in cylindrical boundary condition.

\begin{figure*}[htbp]
  \begin{center}
    \begin{tabular}{r}
      \begin{minipage}{0.5\hsize}
        \begin{center}
          \includegraphics[clip, width=20pc]{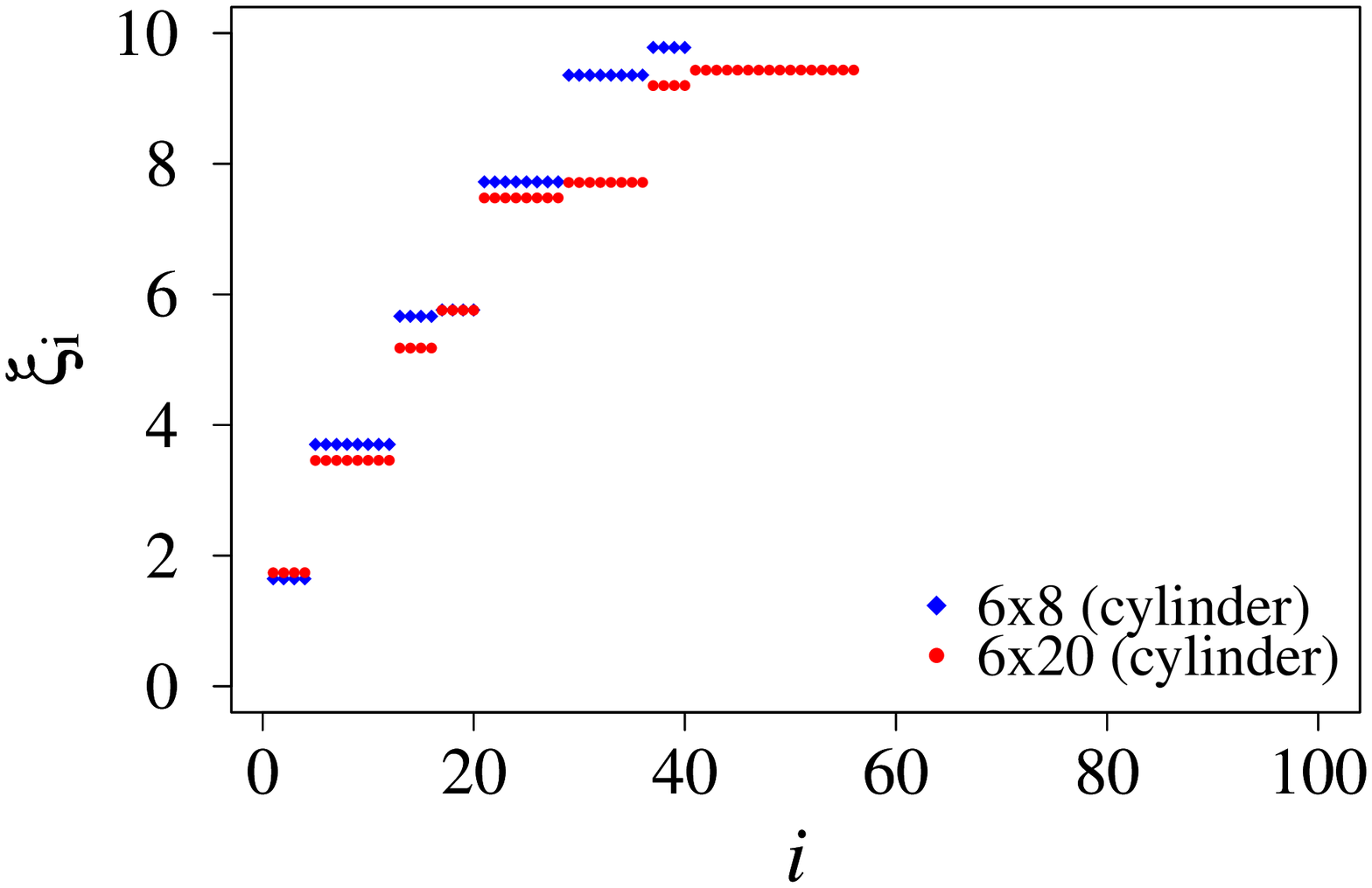}
          \hspace{1cm} (a)
          \vspace{1cm}
        \end{center}
      \end{minipage}
      \begin{minipage}{0.5\hsize}
        \begin{center}
          \includegraphics[clip, width=20pc]{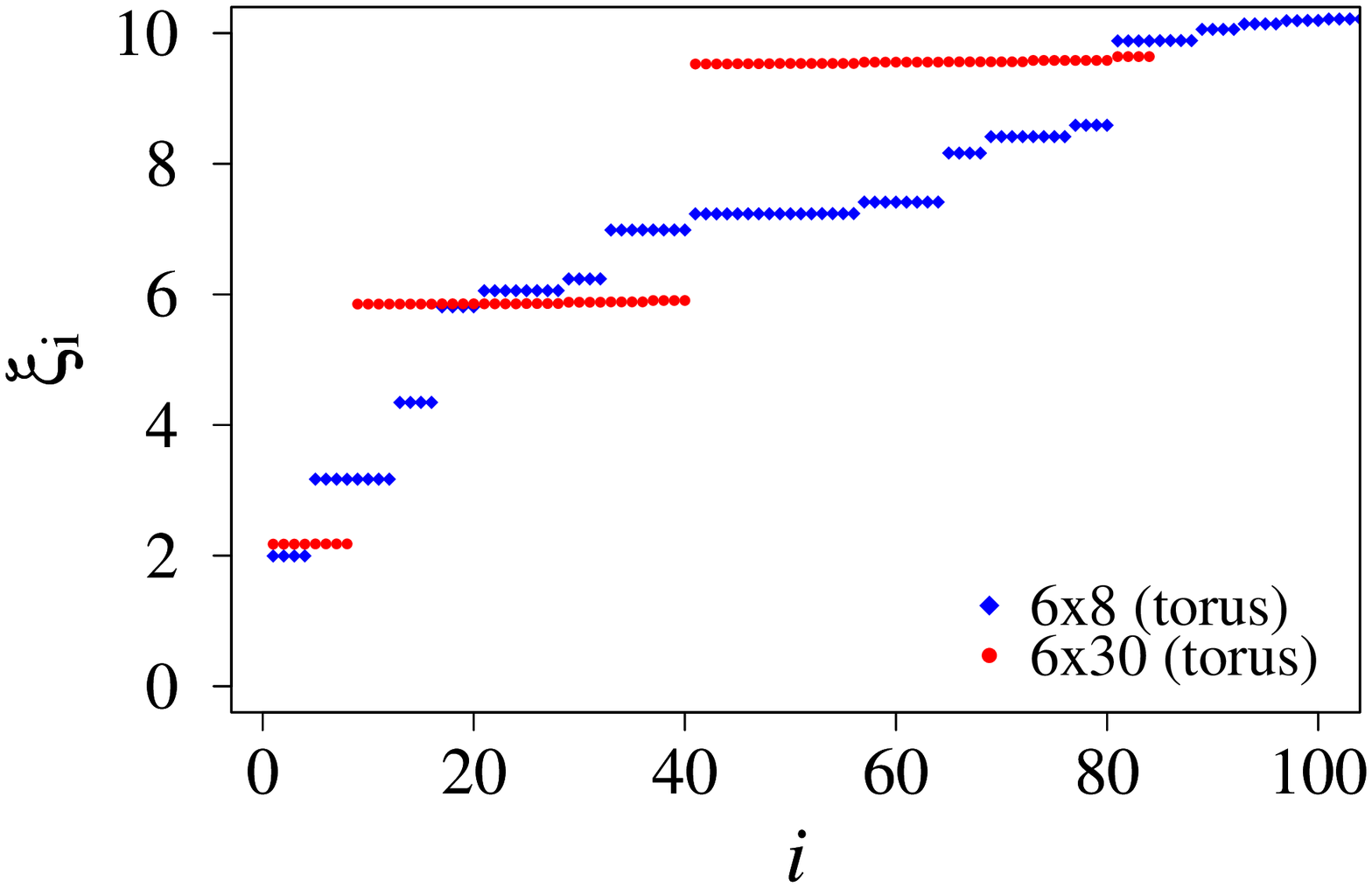}
          \hspace{1cm} (b)
          \vspace{1cm}
        \end{center}
      \end{minipage}
      \\
      \begin{minipage}{0.5\hsize}
        \begin{center}
          \includegraphics[clip, width=20pc]{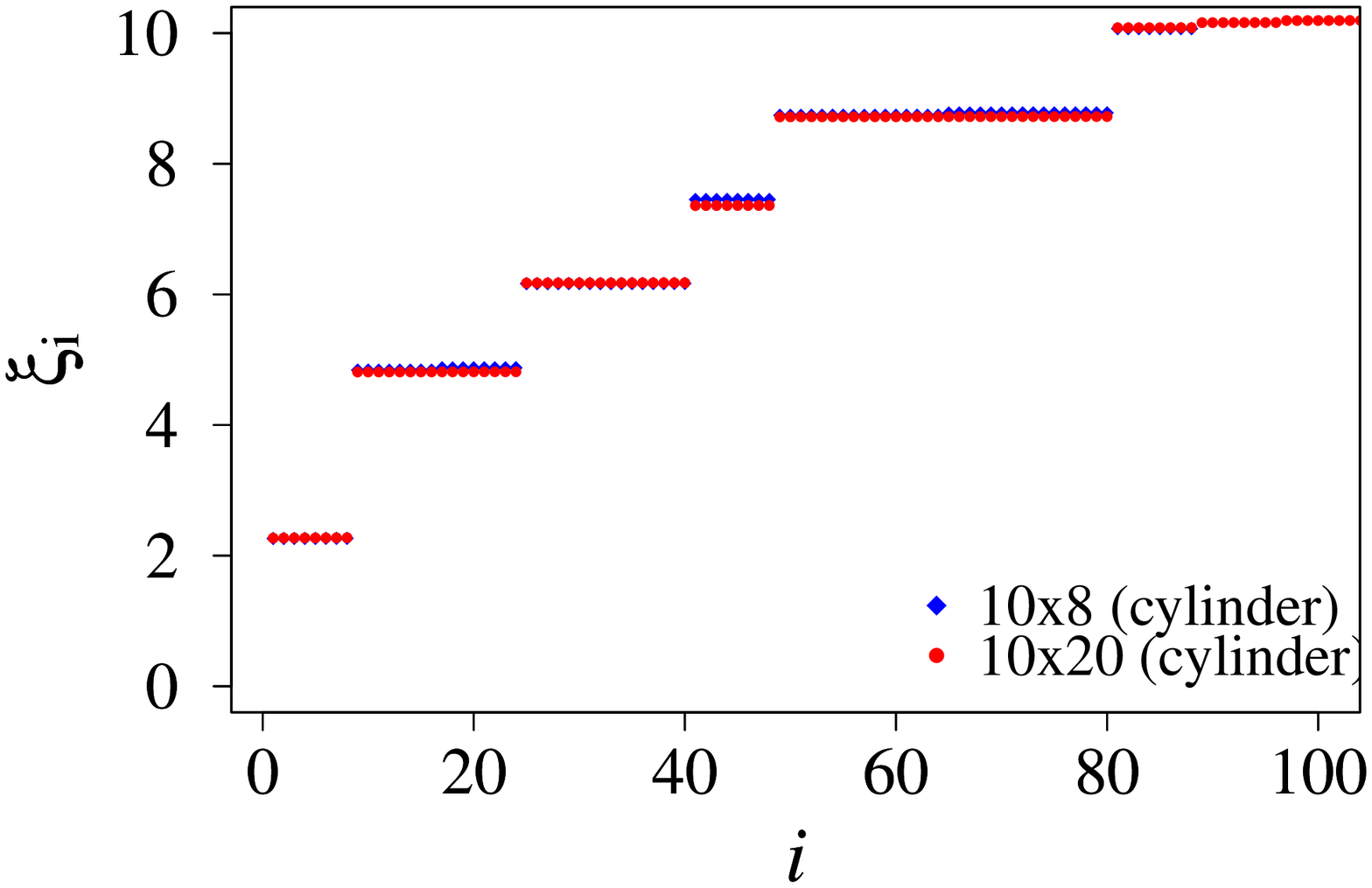}
          \hspace{1cm} (c)
          \vspace{1cm}
        \end{center}
      \end{minipage}
      \begin{minipage}{0.5\hsize}
        \begin{center}
          \includegraphics[clip, width=20pc]{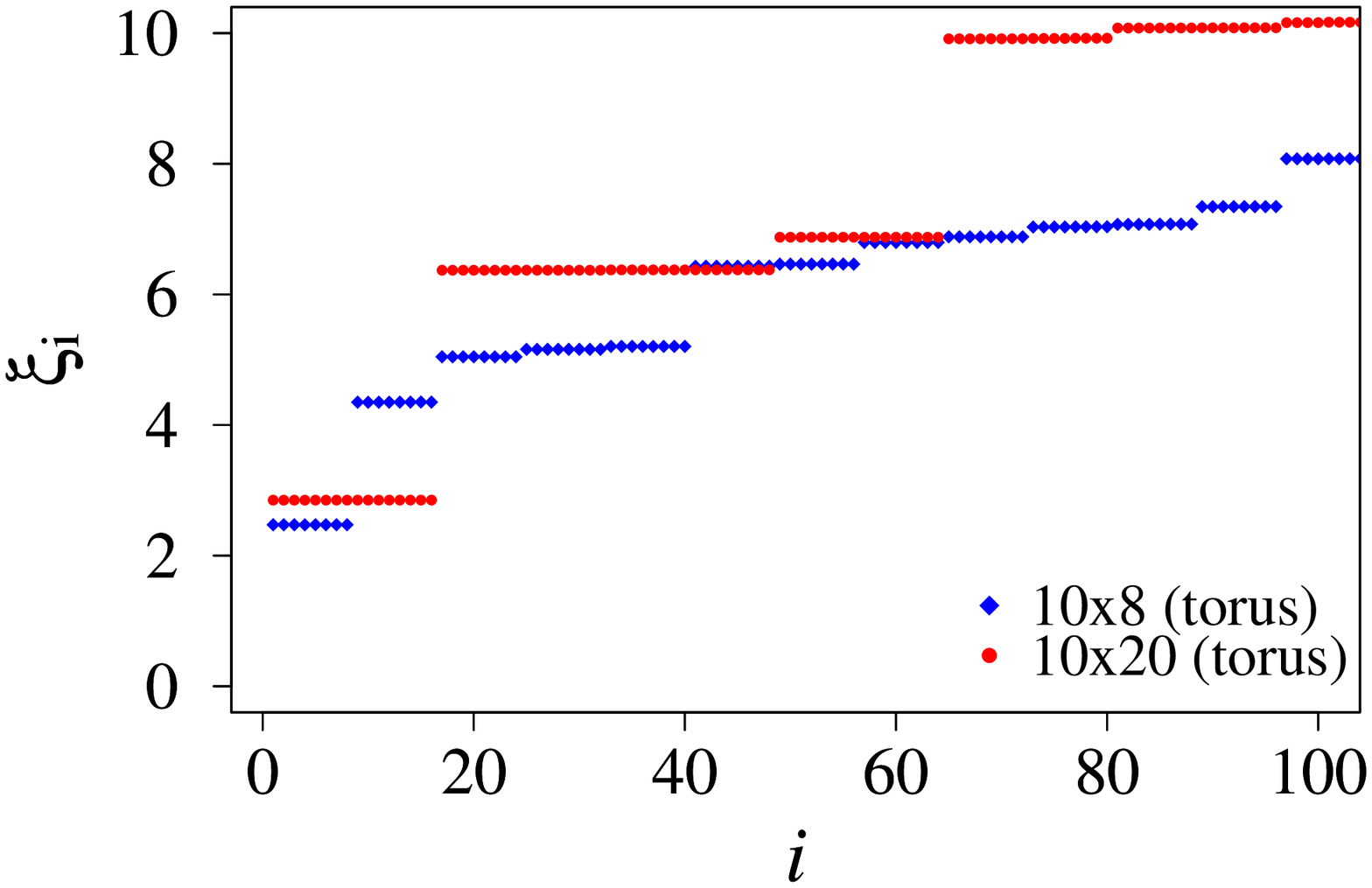}
          \hspace{1cm} (d)
          \vspace{1cm}
        \end{center}
      \end{minipage}

    \end{tabular}
    \caption{The dependence of entanglement spectrum $\xi_i$ of the Kitaev spin-liquid on system size and boundary condition. (a) 6$\times$8-site system (blue rhombuses) and 6$\times$20-site system (red circles) with cylindrical boundary condition, (b) 6$\times$8-site system (blue rhombuses) and 6$\times$30-site system (red circles) with toroidal boundary conditions, (c) 10$\times$8-site system (blue rhombuses) and 10$\times$20-site system (red circles) with cylindrical boundary conditions, and (d)10$\times$8-site system (blue rhombuses) and 10$\times$20-site system (red circles) with toroidal boundary conditions.}
    \label{es_size}
  \end{center}
\end{figure*}

Let us briefly confirm this. Figure~\ref{wilson} shows a 6$\times$8-site system, where the number labels sites on honeycomb lattice and vertical lines denote cutting position when we divide the whole system into two subsystems to calculate entanglement spectrum.
Note that a system with toroidal boundary condition is cut twice at a middle vertical line and a right or left line, while a system with cylindrical boundary condition is cut only once at middle vertical line.
Firstly, we consider the case of cylindrical boundary condition.
Then, the system is divided into A and B parts.
We define the Wilson loops
\begin{align*}
\hat W_1 &= \hat \sigma _{24}^y \hat \sigma _{23}^z \hat  \sigma_{22}^x \hat \sigma _{27}^y \hat \sigma _{26}^z \hat \sigma _{25}^x \\
         &=\hat w_1^A \hat w_1^B , \\
\hat W_2  &=\hat \sigma _{22}^y \hat \sigma _{21}^z \hat \sigma_{20}^x \hat \sigma _{29}^y \hat \sigma _{28}^z \hat \sigma _{27}^x \\
         &=\hat w_2^A \hat w_2^B , 
\end{align*}
where $\sigma _i ^x$, $\sigma _i^y$ and $\sigma _i^z$ are Pauli matrices at $i$-site and
\begin{align*}
\hat w_1^A &=\hat \sigma _{24}^y \hat \sigma _{23}^z \hat \sigma_{22}^x,\\
\hat w_1^B &=\hat \sigma _{27}^y \hat \sigma _{26}^z \hat \sigma _{25}^x,\\
\hat w_2^A &=\hat \sigma _{22}^y \hat \sigma _{21}^z \hat \sigma_{20}^x, \\
\hat w_2^B &=\hat \sigma _{29}^y \hat \sigma _{28}^z \hat \sigma _{27}^x.
\end{align*}
Note that commutation relation
\begin{align}
\left[ \hat W_1,\hat W_2 \right] =0 \label{w1w2}
\end{align}
and anticommutation relations
\begin{align*}
\left \{ \hat w_1^A , \hat w_2 ^A \right \} =0, \\
\left \{ \hat w_1^B , \hat w_2 ^B \right \} =0.
\end{align*}
The ground state is vortex free state, so that the ground state should be an eigenstate of $\hat W_1$ with eigenvalue $+1$:
\begin{align*}
\hat W_1 |\psi \rangle =+|\psi \rangle .
\end{align*}
The ground state can be written as
\begin{align}
|\psi \rangle =& |W_1 = +1\rangle \\
                    =& c_+ |w_1^A =+1, w_1^B=+1 \rangle \nonumber \\
                    &+ c_- | w_1^A=-1, w_1^B=-1 \rangle , \label{psi1}
\end{align}
with
\begin{align*}
\hat w_1^{A,B} |w_1^{A,B} =\pm 1 \rangle = \pm |w_1^{A,B} &=\pm 1 \rangle .
\end{align*}
The eigenstates obey
\begin{align*}
\hat w_2^A |w_1^A  =+1\rangle = | w_1^A=-1 \rangle , \\
\hat w_2^A |w_1^A  =-1\rangle = | w_1^A=+1 \rangle .
\end{align*}
Furthermore, from Eq.~(\ref{w1w2}), the ground state $|\psi \rangle$ is simultaneous eigenstate of $W_1$ and $W_2$, so that $|\psi \rangle$ is also eigenstate of $W_2$:
\begin{align}
\hat W_2 |\psi \rangle =& \hat w_2^A \hat w_2^B | W_1 =+1 \rangle \nonumber \\
                                   =& c_+ \hat w_2^A \hat w_2^B |w_1^A =+1, w_1^B=+1 \rangle \nonumber \nonumber \\
                                   &+ c_- \hat w_2^A \hat w_2^B | w_1^A=-1, w_1^B=-1 \rangle \nonumber \\
                                   =& c_+ |w_1^A =-1, w_1^B=-1 \rangle \nonumber \\
                                   &+ c_- | w_1^A=+1, w_1^B=+1 \rangle \label{psi2} \\
                                   =&+|\psi \rangle. \nonumber 
\end{align}
Therefore, comparing Eq.~(\ref{psi1}) and Eq.~(\ref{psi2}), we obtain
\begin{align*}
c_+=c_- \equiv c
\end{align*}
and
\begin{align*}
|\psi \rangle = c \left( |w_1^A =+1, w_1^B=+1 \rangle + | w_1^A=-1, w_1^B=-1 \rangle \right) .
\end{align*}
The reduced density-matrix of subsystem A reads
\begin{align*}
\rho _A =& {\rm Tr}_B \rho = {\rm Tr} _B |\psi \rangle \langle \psi |  \\
             =& \langle w_1^B=1| \rho |w_1^B=1\rangle + \langle w_1^B=-1|\rho |w_1^B=-1\rangle  \\
             =& c \left( |w_1^A =+1 \rangle \langle w_1^A=+1 | +  |w_1^A =-1 \rangle \langle w_1^A=-1 | \right) \\
             =&c
             \begin{pmatrix}
                  1 & 0 \\
                  0 & 1
              \end{pmatrix}
\end{align*}
Therefore, we find that the eigenvalues of $\rho _A$, i.e., entanglement spectra, are 2-fold degenerate.

It is possible to define the third Wilson loop above $W_2$, which shares the 24-25 (20-29) edge with $W_1$ ($W_2$) loop.
However, the same procedure as (\ref{psi2}) with respect to the third loop will give a result similar to the case of $W_2$. This means no additional state for $\rho _A$, and thus the third loop does not contribute to increasing the number of degeneracy.

Next we consider a system with toroidal boundary condition in a similar way.
In this case, we define additional Wilson loops $W_3$ and $W_4$ that are located on the pink line at the edge of Fig. \ref{wilson}.
These Wilson loops contribute to additional degeneracy of entanglement spectrum, resulting in $2^2$-fold degeneracy of entanglement spectrum.
A similar result with this discussion has been obtained by Yao and Qi,~\cite{yao2010} where the number of degeneracy of entanglement spectrum is $2^{L-1}$ with $L$ being the length of boundary between the A and B subsystems.

Based on the discussion above, we expect that the number of degeneracy in a $10\times L_x$-site system is larger than that in a $6\times L_x$-site system, since the length of boundary between A and B is longer, i.e., the number of Wilson loops defined on honeycomb lattice is larger in the former than in the latter. We confirm this by our DMRG calculations as shown in Fig.~\ref{es_size}, were we keep 700 states in the DMRG block and performed more than 20 sweeps, resulting in truncation error $10^{-10}$ or smaller.

Blue rhombuses and red circles in Fig.~\ref{es_size}(a) show low entanglement levels for cylindrical 6$\times$8-site and 6$\times$20-site systems, respectively.
We find that the levels are at least 4-fold degenerate.
The results for the same system but with toroidal boundary condition are shown in Fig.~\ref{es_size}(b), where in contrast with cylindrical boundary condition, the number of degeneracy strongly depends on the system size along the $x$-axis direction: at least 4-fold degeneracy for 6$\times$8-site system and at least 8-fold degeneracy for 6$\times$30-site system.
We also examined 6$\times$12-site and 6$\times$20-site systems and obtained the same result (not shown). Therefore, we can conclude that 4-fold degeneracy for 6$\times L_x$ with cylindrical boundary condition and 8-fold with toroidal boundary condition as discussed above.

Next, we enlarge system along the $y$-axis direction.
Blue rhombuses and red circles in Fig.~\ref{es_size} (c) show low entanglement levels for cylindrical 10$\times$8-site and 10$\times$20-site systems, respectively.
We find that the levels are at least 8-fold degenerate and thus the degeneracy is doubled as compared with 6$\times L_x$-site system.
The results for the same system but with toroidal boundary condition are shown in Fig.~\ref{es_size}(d), where in contrast with cylindrical boundary condition, the number of degeneracy strongly depends on the system size along the $y$-axis direction: at least 8-fold degeneracy for 10$\times$8-site system and at least 16-fold degeneracy for 10$\times$20-site system.
Therefore, we can conclude that 8-fold degeneracy for 10$\times$$L_x$ with cylindrical boundary condition and 16-fold with toroidal boundary condition. All of these numerical results are consistent with an analytical ones mentioned above.

The ground state of the Kitaev spin-liquid state can be regarded as Majorana fermions coupled with $\mathbb{Z}_2$ gauge field.  The gauge field is, thus, the origin of the degeneracy of entanglement spectrum.
We note that such a gauge fluctuation also affects topological entanglement entropy.\cite{yao2010}

\bibliographystyle{junsrt}

\begin{thebibliography}{99}
\bibitem{kitaev2006} A. Kitaev, Ann. Phys. (NY) {\bf 321}, 2 (2006).
\bibitem{knolle2014} J. Knolle, D. L. Kovrizhin, J. T. Chalker, and R. Moessner, Phys. Rev. Lett. {\bf 112}, 207203 (2014).
\bibitem{baskaran2007} G. Baskaran, S. Mandal, and R. Shankar, Phys. Rev. Lett. {\bf 98}, 247201 (2007).
\bibitem{mandal2011} S. Mandal, S. Bhattacharjee, K. Sengupta, R. Shankar, and G. Baskaran, Phys. Rev. B {\bf 84}, 155121 (2011).
\bibitem{jackeli2009} G. Jackeli and G. Khaliullin, Phys. Rev. Lett. {\bf 102}, 017205 (2009).
\bibitem{chaloupka2010} J. Chaloupka, G. Jackeli, and G. Khaliullin, Phys. Rev. Lett. {\bf 105}, 027204 (2010).
\bibitem{jiang2011} H.-C. Jiang, Z.-C. Gu, X.-L. Qi, and S. Trebst, Phys. Rev. B {\bf 83}, 245104 (2011).
\bibitem{reuther2011} J. Reuther, R. Thomale, and S. Trebst, Phys. Rev. B {\bf 84}, 100406 (2011).
\bibitem{okamoto2013} S. Okamoto, Phys. Rev. B {\bf 87}, 064508 (2013).
\bibitem{schaffer2013} R. Schaffer, S. Bhattacharjee, and Y. B. Kim, Phys. Rev. B {\bf 86}, 224417 (2012).
\bibitem{chaloupka2013} J. Chaloupka, G. Jackeli, and G. Khaliullin, Phys. Rev. Lett. {\bf 110}, 097204 (2013).
\bibitem{price2013} C. Price and N. B. Perkins, Phys. Rev. B {\bf 88}, 024410 (2013).
\bibitem{sela2014} E. Sela, H.-C. Jiang, M. H. Gerlach, and S. Trebst, Phys. Rev. B {\bf 90}, 035113 (2014).
\bibitem{kugel1982} K. Kugel and D. Khomskii, Sov. Phys. Usp. {\bf 25}, 231 (1982).
\bibitem{khaliullin2005} G. Khaliullin, Prog. Theor. Phys. Suppl. {\bf 160}, 155 (2005).
\bibitem{nussinov2013} Z. Nussinov and J. van den Brink, ArXiv e-prints (2013), arXiv:1303.5922 [cond-mat.str-el].\bibitem{choi2012} S. K. Choi, R. Coldea, A. N. Kolmogorov, T. Lancaster, I. I. Mazin, S. J. Blundell, P. G. Radaelli, Y. Singh, P. Gegenwart, K. R. Choi, S.-W. Cheong, P. J. Baker, C. Stock, and J. Taylor, Phys. Rev. Lett. {\bf 108}, 127204 (2012).

\bibitem{ye2012} F. Ye, S. Chi, H. Cao, B. C. Chakoumakos, J. A. Fernandez-Baca, R. Custelcean, T. F. Qi, O. B. Korneta, and G. Cao, Phys. Rev. B {\bf 85}, 180403 (2012).
\bibitem{kimchi2011} I. Kimchi and Y.-Z. You, Phys. Rev. B {\bf 84}, 180407 (2011).
\bibitem{albuquerque2011} A. F. Albuquerque, D. Schwandt, B. Het\'{e}nyi, S. Capponi, M. Mambrini, and A. M. L\"{a}uchli, Phys. Rev. B {\bf 84}, 024406 (2011).
\bibitem{bhattacharjee2012} S. Bhattacharjee, S.-S. Lee, and Y. B Kim, New J. Phys. {\bf 14}, 073015 (2012).
\bibitem{yamaji2014} Y. Yamaji, Y. Nomura, M. Kurita, R. Arita, and M. Imada, Phys. Rev. Lett. {\bf 113}, 107201 (2014).
\bibitem{rau2014} J. G. Rau, E. K.-H. Lee, and H.-Y. Kee, Phys. Rev. Lett. {\bf 112}, 077204 (2014).
\bibitem{rau2014b} J. G. Rau and H.-Y. Kee, ArXiv e-prints (2014), arXiv:1408.4811 [cond-mat.str-el]
\bibitem{sizyuk2014} Y. Sizyuk, C. Price, P. W\"{o}lfle, and N. B. Perkins, ArXiv e-prints (2014), arXiv:1408.3647 [cond-mat.str-el].
\bibitem{kimchi2014} I. Kimchi, R. Coldea, and A. Vishwanath, ArXiv e-prints (2014), arXiv:1408.3640 [cond-mat.str-el].
\bibitem{katukuri2014} V. M. Katukuri, S. Nishimoto, V. Yushankhai, A. Stoyanova, H. Kandpal, S. Choi, R. Coldea, I. Rousochatzakis, L. Hozoi, and J. van den Brink, New J. Phys. {\bf 16}, 013056 (2014).
\bibitem{reuther2014} J. Reuther, R. Thomale, and S. Rachel, Phys. Rev. B {\bf 90}, 100405(R) (2014).
\bibitem{white1992} S. R. White, Phys. Rev. Lett. {\bf 69}, 2863 (1992).
\bibitem{schollwock2005} U. Schollw\"{o}ck, Rev. Mod. Phys. {\bf 77}, 259 (2005).
\bibitem{li2008} H. Li and F. D. M. Haldane, Phys. Rev. Lett. {\bf 101}, 010504 (2008).
\bibitem{holzhey1994} C. Holzhey, F. Larsen, and F. Wilczek, Nucl. Phys. B {\bf 424}, 443 (1994).
\bibitem{calabrese2004} P. Calabrese and J. Cardy, J. Stat. Mech. (2004) P06002.
\bibitem{yao2010} H. Yao and X.-L Qi, Phys. Rev. Lett. {\bf 105}, 080501 (2010).
\bibitem{chandran2014} A. Chandran, V. Khemani, and S. L. Sondhi, Phys. Rev. Lett. {\bf 113}, 060501 (2014).
\bibitem{furukawa2006} S. Furukawa, G. Misguich, and M. Oshikawa, Phys. Rev. Lett. {\bf 96}, 047211 (2006).
\bibitem{henley2014} C. L. Henley and H. J. Changlani, ArXiv e-prints (2014), arXiv:1407.4189 [cond-mat.stat-phys].

\bibitem{regnault2009} N. Regnault, B. A. Bernevig, and F. D. M. Haldane, Phys. Rev. Lett. {\bf 103}, 016801 (2009).
\bibitem{lauchli2010} A. M. L\"{a}uchli, E. J. Bergholtz, J. Suorsa, and M. Haque, Phys. Rev. Lett. {\bf 104}, 156404 (2010).
\bibitem{thomale2010may} R. Thomale, A. Sterdyniak, N. Regnault, and B. A. Bernevig, Phys. Rev. Lett. {\bf 104}, 180502 (2010).
\bibitem{turner2010} A. M. Turner, Y. Zhang, and A. Vishwanath, Phys. Rev. B {\bf 82}, 241102 {2010}.
\bibitem{fidkowski2010} L. Fidkowski, Phys. Rev. Lett. {\bf 104}, 130502 (2010).
\bibitem{thomale2010sep} R. Thomale, D. P. Arovas, and B. A. Bernevig, Phys. Rev. Lett. {\bf 105}, 116805 (2010).
\bibitem{lundgren2012} R. Lundgren, V. Chua, and G. A. Fiete, Phys. Rev. B {\bf 86}, 224422 (2012).
\bibitem{lepori2013} L. Lepori, G. De Chiara, and A. Sanpera, Phys. Rev. B {\bf 87}, 235107 (2013).
\bibitem{giampaolo2013} S. M. Giampaolo, S. Montangero, F. Dell'Anno, S. De Siena, and F. Illuminati, Phys. Rev. B {\bf 88}, 125142 (2013).
\bibitem{james2013} A. J. A. James and R. M. Konik, Phys. Rev. B {\bf 87}, 241103 (2013).
\bibitem{fradkin} E. Fradkin, Field Theories of Condensed Matter Physics, Cambridge University Press, Cambridge (2013).
\bibitem{morita} S. Morita, private communications.
\end{thebibliography}

 \end{document}